\documentclass[structabstract]{aa}
\usepackage{graphicx}
\usepackage{txfonts}
\begin{document}
   \title{Chemical study of intermediate-mass (IM) Class 0 protostars.}
   \subtitle{CO depletion and N$_2$H$^+$ deuteration}

   \author{T. Alonso-Albi
          \inst{1}
          \and
        A. Fuente\inst{1}
          \and
	N. Crimier
	 \inst{2}
	\and
          P. Caselli\inst{3}
        \and
         C. Ceccarelli\inst{2}
	\and
         D. Johnstone\inst{4,5}
        \and 
        P. Planesas\inst{1,6}
        \and
        J.R. Rizzo\inst{7}
        \and
       F. Wyrowski\inst{8}
        \and
        M. Tafalla\inst{1}
        \and
        B. Lefloch\inst{2}
       \and
       S. Maret\inst{2}
	\and
       C. Dominik\inst{9}
   }

   \institute{Observatorio Astron\'omico Nacional (OAN,IGN), Apdo 112, E-28803 Alcal\'a de Henares, Spain
	\email{t.alonso@oan.es}
             \and
             Laboratoire d'Astrophysique Observatoire de Grenoble, BP 53, F-38041 Grenoble C\'edex 9, France
	     \and
             School of Physics \& Astronomy, E.C. Stoner Building, The University of Leeds, Leeds LS2 9JT, UK 
            \and
     	     Department of Physics \& Astronomy, University of Victoria, Victoria, BC, V8P 1A1, Canada
	    \and
            National Research Council of Canada, Herzberg Institute of Astrophysics, 5071 West Saanich Road,
            Victoria, BC, V9E 2E7, Canada
           \and
        Joint ALMA Observatory, El Golf 40, Las Condes, Santiago, Chile
	\and
           Centro de Astrobiolog\'{\i}a (CSIC/INTA), Laboratory of Molecular Astrophysics, Ctra. Ajalvir km. 4, E-28850, Torrej\'{o}n de Ardoz, Spain
          \and
          Max-Planck-Institut f\"ur Radioastronomie, Auf dem H\"ugel 69, D-53121 Bonn, Germany
       \and
      Anton Pannekoek Astronomical Institute, University of Amsterdam, P.O. Box 94249, 1090 GE Amsterdam, The
Netherlands
}

   \date{Received February 2, 2010; accepted April 4, 2010}

  \abstract
   {}
   {
We are carrying out a physical and chemical study of the protostellar envelopes
in a representative sample of IM Class 0 protostars. In our first paper we 
determined the physical structure (density-temperature radial profiles) of the
protostellar envelopes. Here, we study the CO depletion and N$_2$H$^+$ deuteration.}
   {We observed the millimeter lines of C$^{18}$O, C$^{17}$O,
N$_2$H$^+$ and N$_2$D$^+$ towards the protostars using the IRAM 30m telescope. Based on these
observations, we derived the C$^{18}$O, N$_2$H$^+$ and N$_2$D$^+$ radial
abundance profiles across their envelopes using a radiative transfer
code. In addition, we modeled the chemistry of the protostellar
envelopes.}
   {All the C$^{18}$O 1$\rightarrow$0 maps are well fit when assuming that
the C$^{18}$O abundance decreases inwards within the protostellar envelope
until the gas and dust reach the CO evaporation temperature,
$\approx$20--25~K, where the CO is released back to the gas phase. 
The N$_2$H$^+$ deuterium fractionation in Class 0 IMs is
[N$_2$D$^+$]/[N$_2$H$^+$]=0.005--0.014,  two orders of magnitude higher than the
elemental [D/H] value in the interstellar medium, but a factor of 10 lower
than in prestellar clumps. 
Chemical models account for the C$^{18}$O and N$_2$H$^+$ observations if we assume the CO abundance is a factor
of $\sim$2 lower than the canonical value in the inner envelope. This could be the consequence of the CO being converted into
CH$_3$OH on the grain surfaces prior to the evaporation and/or the photodissociation of CO by the stellar
UV radiation. The deuterium fractionation 
is not fitted by chemical models.
This discrepancy is very likely caused by the simplicity of our model that assumes spherical geometry and neglects important
phenomena like the effect of bipolar outflows and UV radiation from the
star. More important, the deuterium fractionation is dependent on the ortho-to-para H$_2$ ratio, which is not likely to reach the 
steady-state value in the dynamical time scales of these protostars.  }
{}

   \maketitle
%
%

\section{Introduction}

Intermediate-mass young stellar objects (IMs) 
share many characteristics
with high-mass stars (clustering, PDRs) but their study presents an important 
advantage: many are located closer to the Sun (d $\leq$ 1 kpc) and 
in less complex regions than massive star-forming regions. 
On the other hand, they are also important for understanding planet formation since Herbig Ae stars are the precursors of Vega-type systems. 
Despite this, IMs have been studied very little so far. A few works on HAEBE stars have been carried out
at millimeter wavelengths (Fuente et al. 1998, 2002; Henning et al. 1998), but almost nothing has been done 
for their precursors, the Class 0 IM objects.

Chemistry has been successfully used to determine the physical structure and 
investigate the formation and evolution of low-mass YSOs. Chemical diagnostics have also been shown to be good indicators of
the protostellar evolution in these objects
(see e.g. Maret et al. 2004, J{\o}rgensen et al. 2005). 
However, very few works deal with IMs.
Fuente et al. (2005a) present a chemical study of the
envelopes of the Class 0 IM protostar NGC~7129--FIRS~2 and the young
Herbig Be star LkH$\alpha$~234. They find that the changes 
in the physical conditions of the envelope during its evolution from the Class 0 to the Class I
stage (the envelope is dispersed and warmed up) strongly influence the molecular chemistry.
The Class 0 object NGC~7129--FIRS~2 presented evidence of H$^{13}$CO$^+$ depletion. 
Moreover, the deuterium fractionation, measured as the DCO$^+$/H$^{13}$CO$^+$
ratio, decreases by a factor of 4 from the Class 0 to the Herbig Be star, very likely
owing to the increase in the kinetic temperature. Regarding the abundance of complex molecules, 
the beam-averaged abundances of CH$_3$OH and H$_2$CO increase
from the Class 0 to the Herbig Be star. A hot core was also detected
in NGC~7129--FIRS~2 (Fuente et al. 2005a,b). Although two objects are not enough to establish
firm conclusions, these pioneering results suggest that chemistry is
also a good indicator of the evolution of IMs. 

We are carrying out a chemical study of a representative sample of IM Class 0 YSOs 
This is the first systematic chemical study of IM Class 0  objects that has been carried out so far. 
Some properties like the temperature of the protostellar envelope
and the clustering degree depend on the final stellar mass, so the results for
low-mass stars cannot be directly extrapolated to intermediate-mass objects.
In the first paper (Crimier et al. 2010, hereafter C10), we determined the physical structure (density-temperature radial profiles) by modeling the
dust continuum emission.
We now present the observations of the millimeter lines of C$^{18}$O, C$^{17}$O,  
N$_2$H$^+$, and N$_2$D$^+$ in the same sample. Our goal is to investigate the CO depletion and N$_2$H$^+$ 
deuteration in these Class 0 YSOs. For comparison, we also include
2 Class I objects, LkH$\alpha$~234, and S140. 

\section{Observational strategy}


Our selection was made to have a representative sample of
Class 0 IM YSOs, including targets with different luminosities 
(40 -- 10$^3$~L$_\odot$) and evolutionary stages. An important complication
in the study of massive stars is that they are located in complex regions
and are therefore difficult to model. 
The targets in this sample were chosen to lie preferentially in isolated areas with respect to the 30m telescope beam.
We also selected sources for which continuum maps at
submillimeter and/or millimeter wavelengths are available in order
to be able to model their envelopes.
The list of sources and their coordinates are shown in Table 1.
To provide a comparison with Class
I sources, we added S140 and LkH$\alpha$~234 to the sample.


Most of the observations reported here were carried out with
the IRAM 30m telescope at Pico de Veleta (Spain)
during three different observing periods in June 2004 in position switching mode. 
Our strategy was to first make long integration 
single-pointing observations towards the star position and
then to make 96$''$$\,\times\ $96$''$ maps around the center position. 
The maps were sampled with a spacing of 12$''$ in the inner 48$''$ regions
and 24$''$  outside. The only exceptions were L1641 S3 MMS1 and S140.
In L1641 S3 MMS1, we only observed a radial strip in
C$^{17}$O and C$^{18}$O. We did not observe  C$^{17}$O and C$^{18}$O maps in
S140. A summary of the observations for each source is shown in Table 1, and the list of observed lines and the telescope characteristics is shown in 
Table 2.
During the observations,  lines of the same species
were observed simultaneously using the multireceiver capability of the 30m telescope.
In this way, we minimized relative pointing and calibration errors. As backends we used in parallel 
an autocorrelator split into several parts providing a spectral resolution that was always
better than $\sim$78 kHz and a 1~MHz-channel-filter-bank.
Examples of the single-pointing observations are shown in Figs. 1 and 2. 
The intensity scale is the main brightness temperature.

Observations of the N$_2$H$^+$ J=4$\rightarrow$3 line (freq=372.6725~GHz)
were carried out towards Cep E--mm, IC~1396~N, NGC~7129--FIRS~2, and L1641 S3 MMS1 using the 
JCMT telescope at the Mauna Kea (Hawaii). In all these sources, we carried out small maps
of 75$''$$\times$75$''$ with a spacing of 25$''$, but the emission was only detected towards 
the star position (see Fig. 3). All the lines were observed with a spectral resolution of 0.488~MHz.  The spectra of the J=4$\rightarrow$3 line are
shown in Fig. 3.

In this paper, we also use the C$^{18}$O, N$_2$H$^+$ and N$_2$D$^+$ data
towards NGC 7129-FIRS 2 and LkH$\alpha$~234, which has already been published by Fuente et al. (2005b).

\begin{table*}
\caption{Selected sample}
\begin{tabular}{lccccc}\\  \hline \hline
\multicolumn{1}{c}{Object} &  \multicolumn{1}{c}{RA(2000)} &
\multicolumn{1}{c}{Dec(2000)} & \multicolumn{1}{c}{Lum. (L$_\odot$)} & 
\multicolumn{1}{c}{d}  &  \multicolumn{1}{c}{} \\ \hline     
Serpens-FIRS 1    & 18:29:49.6  &   +01:15:20.6 &   33    & 230  & Maps in C$^{17}$O,C$^{18}$O,N$_2$H$^+$ and
N$_2$D$^+$    \\
Cep E-mm          & 23:03:13.1  &   +61:42:26.0 &   100    & 730  & Maps in C$^{17}$O,C$^{18}$O,N$_2$H$^+$ and
N$_2$D$^+$    \\
L1641 S3 MMS 1    & 05:39:55.9  &   -07:30:28.0 &  67    & 500  & Strip in C$^{17}$O and C$^{18}$O,
maps in N$_2$H$^+$ and
N$_2$D$^+$ \\
IC1396N           & 21:40:41.7  &   +58:16:12.8 &  150    & 750  & Maps in C$^{17}$O,C$^{18}$O,N$_2$H$^+$ and
N$_2$D$^+$ \\     
CB3               & 00:28:42.7  &   +56:42:07.1 &  1000    & 2500 & Maps in C$^{17}$O,C$^{18}$O,N$_2$H$^+$ and
N$_2$D$^+$  \\
OMC2-FIR4        &  05:35:26.7  &  -05:10:00.5  &   1000  &  450 & Maps in C$^{17}$O,C$^{18}$O,N$_2$H$^+$ and
N$_2$D$^+$   \\  
NGC 7129--FIRS 2 & 21:43:01.7 & +66:03:23.6 & 500 & 1250 & Strip in C$^{17}$O, maps in C$^{18}$O,N$_2$H$^+$ and
N$_2$D$^+$  \\
\hline
S140             & 22:19:18. 1 &  +63:18:54.6 &  10$^4$ &  910 &    N$_2$H$^+$ and N$_2$D$^+$ maps    \\
LkH$\alpha$ 234 & 21:43:06.8 & +66:06:54.4   & 500 & 1250 &  Maps in C$^{18}$O,N$_2$H$^+$ and
N$_2$D$^+$ \\
\hline \hline
\end{tabular}
\end{table*}

\begin{table}[t]
\caption{Description of the observations}
\begin{tabular}{lr ccc} \\  \hline \hline
\multicolumn{1}{c}{Line} &       
\multicolumn{1}{c}{Freq. (MHz)} &
\multicolumn{1}{c}{HPBW} &
\multicolumn{1}{c}{$\eta_{MB}$} &
\multicolumn{1}{c}{Tel} 
  \\ \hline
C$^{18}$O     1$\rightarrow$0   &  109782.2   &   22$''$      &  0.75 & IRAM  \\
C$^{18}$O     2$\rightarrow$1   &  219560.3   &   11$''$      &  0.55 & IRAM  \\
C$^{17}$O     1$\rightarrow$0   &  112359.3   &   22$''$      &  0.74 & IRAM  \\
C$^{17}$O     2$\rightarrow$1   &  224714.4   &   11$''$      &  0.54 & IRAM  \\
N$_2$H$^+$    1$\rightarrow$0   &   93173.2   &   26$''$      &  0.77 & IRAM  \\
N$_2$H$^+$    4$\rightarrow$3   &  372672.5   &   13$''$      &  0.63 & JCMT  \\ 
N$_2$D$^+$    2$\rightarrow$1   &  154217.1   &   16$''$      &  0.67 & IRAM  \\ 
N$_2$D$^+$    3$\rightarrow$2   &  231321.7   &   11$''$      &  0.52 & IRAM  \\
\hline \hline
\end{tabular}
\end{table}

\setlength\unitlength{1cm}
\begin{figure*}
\vspace{8cm}
\includegraphics{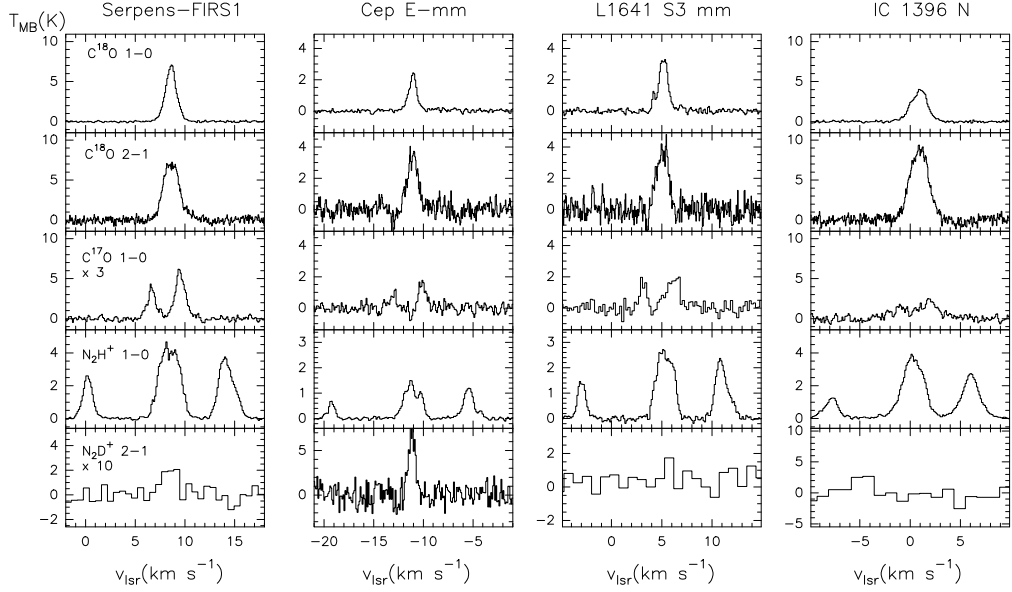}
\caption{Examples of the single-pointing observations towards the Serpens--FIRS 1, Cep E--mm, L1641~S3~MMS1, and IC~1396~N. All the spectra were observed with the 30m telescope. The intensity scale is the main brightness temperature.}
\end{figure*}

\setlength\unitlength{1cm}
\begin{figure*}
\vspace{9.0cm}
\includegraphics{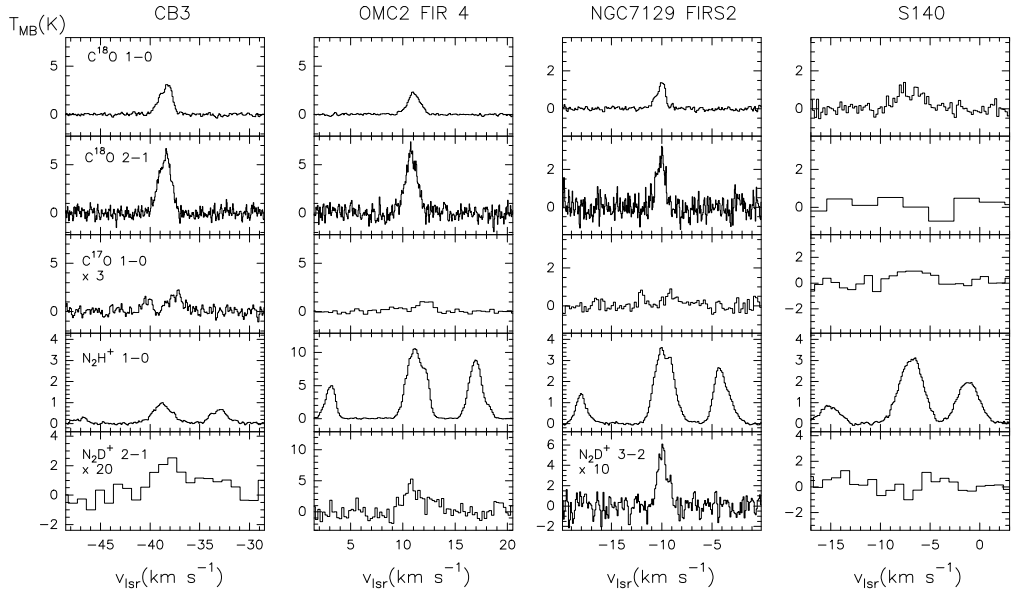}
\caption{The same as Fig. 1 for CB3, OMC2~FIR~4, NGC~7129--FIRS~2, and S140.}
\end{figure*}

\section{Results}
The spectra of the  C$^{18}$O 1$\rightarrow$0, C$^{17}$O 1$\rightarrow$0,
N$_2$H$^+$ 1$\rightarrow$0 and N$_2$D$^+$ 2$\rightarrow$1 lines towards the star
position are shown in Figs. 1 and 2. The integrated intensity maps of the 
C$^{18}$O 1$\rightarrow$0, N$_2$H$^+$ 1$\rightarrow$0,
and N$_2$D$^+$ 2$\rightarrow$1 lines 
are shown in Figs.\ 4 (Class 0) and 5 (Class I).
In all the Class 0 sources, the
emission of the N$_2$H$^+$ 1$\rightarrow$0 line is compact 
and peaks towards the star position, revealing a 
dense core around the star. However, towards the Class I sources, the
peaks of the N$_2$H$^+$ emission are offset from the far-IR source.
For
the Herbig Be star LkH$\alpha$ 234, the emission peak of
the N$_2$H$^+$ 1$\rightarrow$0 line is located
at an offset (-6$"$,18$"$) from the star position (see Fuente et al. 2005a
and Fig.\ 5). In S140, the N$_2$H$^+$ emission peaks in an arc-shaped feature 
surrounding the sources IRS 1, 2 and 3. 
In Class I sources, either  the N$_2$H$^+$ is destroyed by the evaporated  
CO or the outflow, or the UV radiation from the star has
already disrupted the parent core.

The emission in the C$^{18}$O 1$\rightarrow$0 line usually presents a different morphology
from that of N$_2$H$^+$ 1$\rightarrow$0 line. In CB3 and IC~1396N, the emission in the C$^{18}$O 
line has  an elongated shape, much more extended than that of N$_2$H$^+$ (see Fig. 4). 
In OMC2~FIR~4 and NGC~7129-FIRS 2, the emission of the
C$^{18}$O line surrounds the star position instead
of having a maximum towards it (see also
Fuente et al. 2005a). Only in the low-luminosity sources Serpens-FIRS 1 and Cep~E-mm does the
emission from the C$^{17}$O and C$^{18}$O lines peak at
the star position and present a morphology similar to that of
the N$_2$H$^+$ emission. 

For the sources in which the signal-to-noise ratio of the N$_2$D$^+$ map
is high enough, Serpens-FIRS~1 and NGC~7129 FIRS~2, we compared
the N$_2$D$^+$ 2$\rightarrow$1 and N$_2$H$^+$ 1$\rightarrow$0 maps. 
In Serpens-FIRS~1, the N$_2$D$^+$ 2$\rightarrow$1 emission does not peak towards the star, but does in
the case of NGC7129-FIRS~2. 

In Fig.\ 6, we show the mean integrated intensity emission of the C$^{18}$O 1$\rightarrow$0, N$_2$H$^+$ 1$\rightarrow$0,
and N$_2$D$^+$ 2$\rightarrow$1 lines in concentric rings around the protostar.
In all the sources, the emission of the N$_2$H$^+$ 1$\rightarrow$0 line decreases outwards from the
star. The radial profiles of the C$^{18}$O 1$\rightarrow$0 line, however, greatly differ from one source to the next.
The presumably youngest sources, OMC2~FIR~4 and NGC~7129--FIRS~2, show a flat profile. 
This is the expected picture when the CO abundance decreases towards the 
center, mainly because the molecules are frozen onto the grain surfaces.
In Serpens-FIRS 1, Cep E-mm, IC~1396~N, and CB3, the C$^{18}$O 1$\rightarrow$0 emission decrease with the distance 
from the star, similar to N$_2$H$^+$ but with a less steep profile.
Intense emission of the C$^{18}$O 1$\rightarrow$0 line is detected at the
edges of the protostar envelopes, which shows that a lower density envelope also contributes to the emission of this
line. This envelope contribution is especially important for Serpens-FIRS~1. In fact, we must model the envelope contribution in order to fit the C$^{18}$O 1$\rightarrow$0 emission from this protostellar core (see Appendix A1).
One could interpret the different radial profiles of the C$^{18}$O 1$\rightarrow$0 emission as an evolutionary
trend, with the youngest protostars having flat profiles while the oldest have steep profiles.
Given the complexity of these regions, however, one should be cautious and use multiple evolutionary tracers to define relative age.
For instance, a lack of C$^{18}$O emission towards the star position could come from CO depletion in the case of a
very young object or to photodissociation for a borderline Class 0/I.
The radial profiles of the N$_2$D$^+$ 2$\rightarrow$1 (3$\rightarrow$2 for NGC~7129--FIRS~2) are
more uncertain because of the low S/N ratio of the maps.

\setlength\unitlength{1cm}
\begin{figure}
\vspace{6cm}
\includegraphics{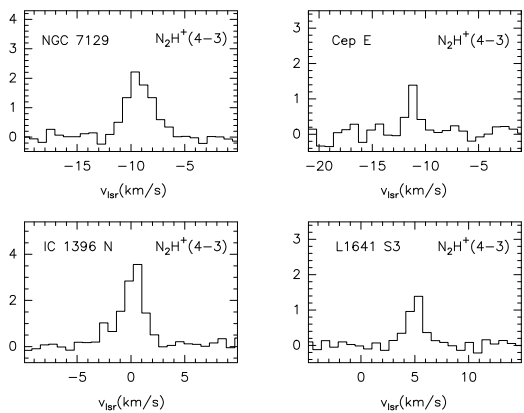}
\caption{Spectra of the N$_2$H$^+$ 4$\rightarrow$3 line towards NGC~7129--FIRS~2, Cep~E-mm, IC~1396~N, and L1641~S3~MMS1.}
\end{figure}

\setlength\unitlength{1cm}
\begin{figure*}
\vspace{21.5cm}
\includegraphics{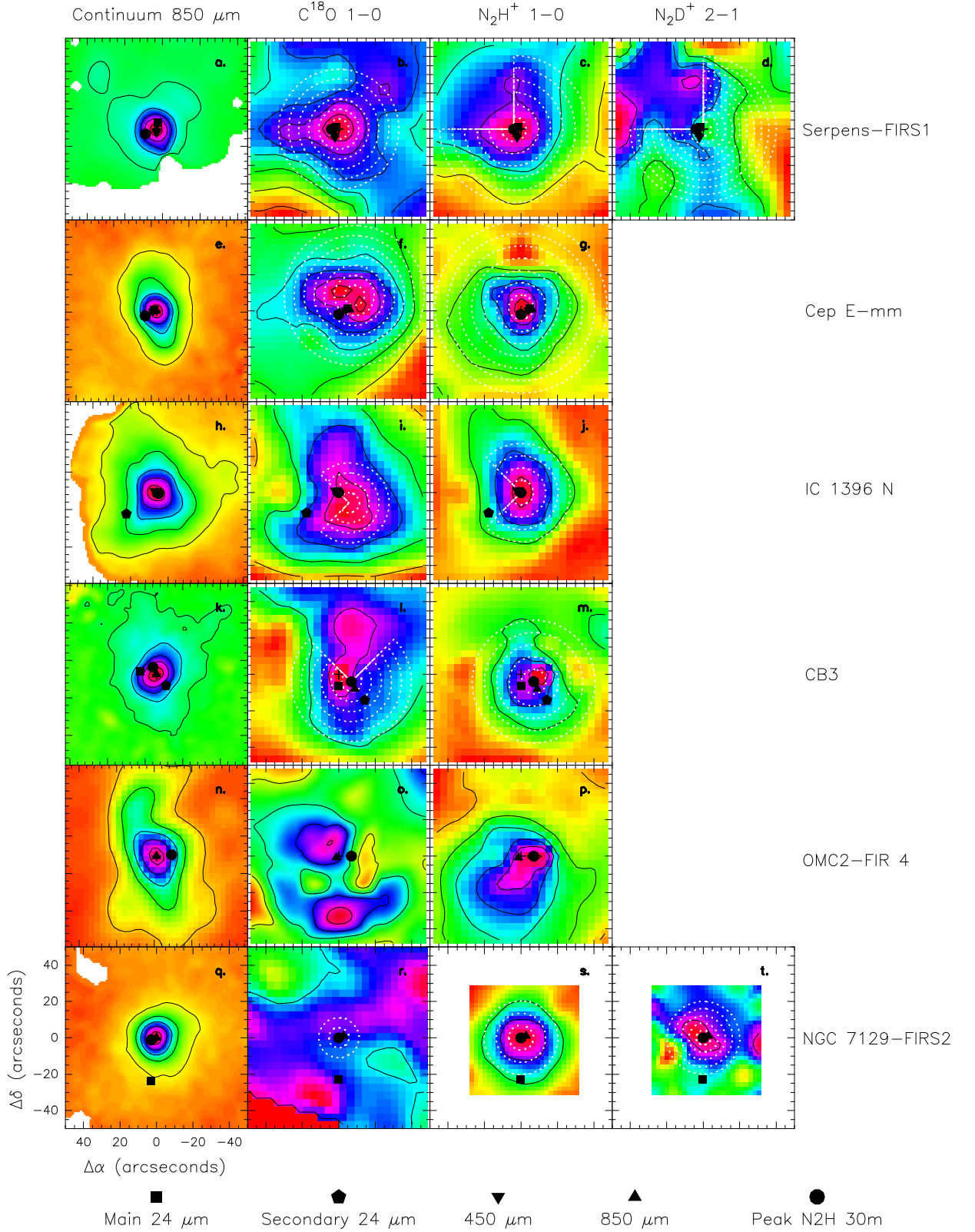}
\caption{Continuum maps at 850~$\mu$m from C10 (left column), and integrated intensity maps of the C$^{18}$O 1$\rightarrow$0, N$_2$H$^+$ 1$\rightarrow$0 and N$_2$D$^+$ 2$\rightarrow$1 (3$\rightarrow$2 in the case of NGC~7129--FIRS~2) lines towards 
the observed Class 0 IMs. In each panel, we have drawn the area and the number of rings considered in our model fitting. The emission peak at several wavelengths is shown with different marks. Contour levels starts at 3$\sigma$ level and are 
{a.} 0.5, 1.6, 2.7, 3.8, 4.9, and 5.4 Jy/beam; 
{b.} 1 to 9~K~km~s$^{-1}$ by steps of 1~K~km~s$^{-1}$; 
{c.} 3 to 9~K~km~s$^{-1}$ by steps of 3~K~km~s$^{-1}$; 
{d.} 0.5 to 1.5~K~km~s$^{-1}$ by steps of 0.5~K~km~s$^{-1}$; 
{e.} 0.16, 0.49, 0.81, 1.14, and 1.46 Jy/beam;
{f.} 0.5 to 3~K~km~s$^{-1}$ by steps of 0.5~K~km~s$^{-1}$; 
{g.} 0.5 to 2.5~K~km~s$^{-1}$ by steps of 0.5~K~km~s$^{-1}$; 
{h.} 0.3, 0.9, 1.5, 2.1, and 2.8 Jy/beam;
{i.} 2.0 to 8.0~K~km~s$^{-1}$ by steps of 2.0~K~km~s$^{-1}$; 
{j.} 3.0 to 9.0~K~km~s$^{-1}$ by steps of 3.0~K~km~s$^{-1}$;
{k.} 0.07, 0.20, 0.33, 0.46, and 0.59 Jy/beam;
{l.} 2.0, 4.0~K~km~s$^{-1}$; 
{m.} 1.0 to 3.0~K~km~s$^{-1}$ by steps of 1.0~K~km~s$^{-1}$;
{n.} 1.0, 2.5, 4.0, 5.5, and 7.0 Jy/beam; 
{o.} 0.5 to 3.5~K~km~s$^{-1}$ by steps of 0.5~K~km~s$^{-1}$;
{p.} 4.0 to 24.0~K~km~s$^{-1}$ by steps of 4.0~K~km~s$^{-1}$;
{q.} 2.1, 6.3, 10.6, 14.8, and 19.0 Jy/beam; 
{r.} 2.0~K~km~s$^{-1}$;
{s.} 3.0, 6.0~K~km~s$^{-1}$;
{t.} 0.5~K~km~s$^{-1}$.
}
\end{figure*}

\setlength\unitlength{1cm}
\begin{figure*}
\vspace{11cm}
\includegraphics{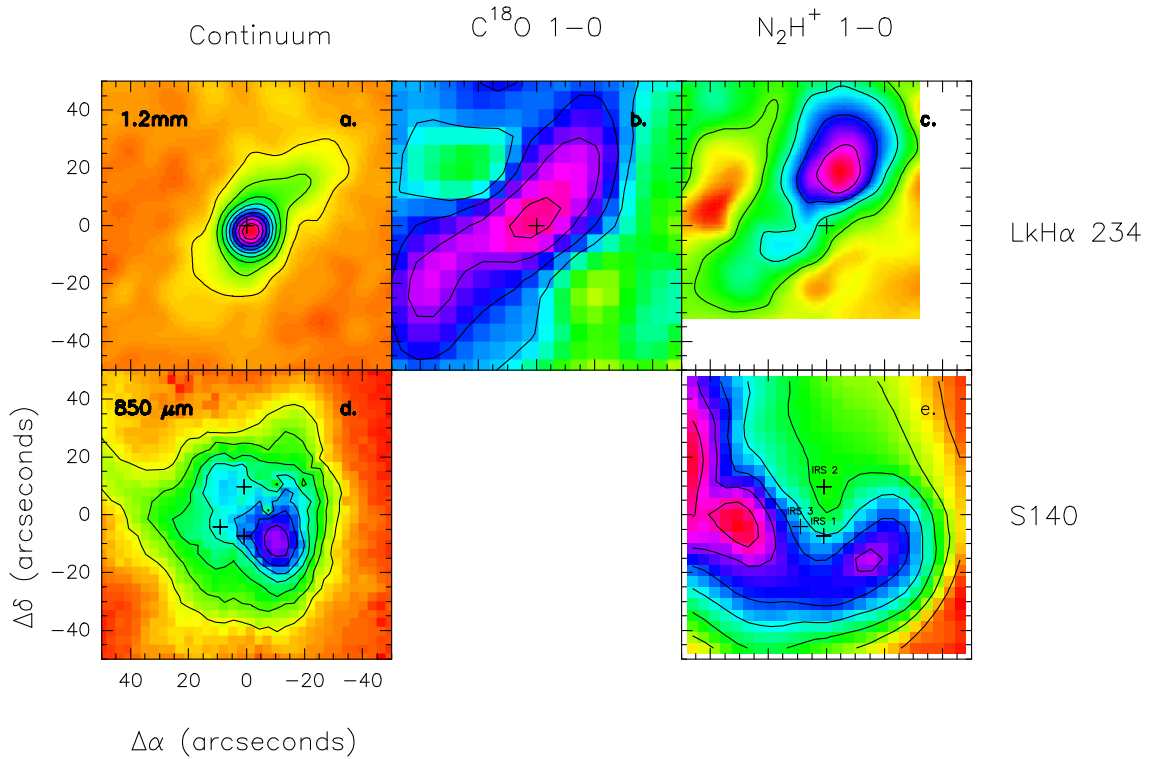}
\caption{Integrated intensity maps of the C$^{18}$O 1$\rightarrow$0 and N$_2$H$^+$ 1$\rightarrow$0 lines towards 
LkH$\alpha$~234 and S140. Contour levels are: 
{\bf a.} 0.07, 0.15, 0.23, 0.31, 0.39, 0.46, 0.54, 0.62, and 0.69 Jy/beam;
{\bf b.} 2 to 5~K~km~s$^{-1}$ by steps of 1~K~km~s$^{-1}$; 
{\bf c.} 1 to 5~K~km~s$^{-1}$ in steps of 1~K~km~s$^{-1}$; 
{\bf d.} 1 to 8.0~Jy/beam in steps of 1~Jy/beam ; 
{\bf e.} 3 to 21~K~km~s$^{-1}$ in steps of 3~K~km~s$^{-1}$.
}
\end{figure*}

\setlength\unitlength{1cm}
\begin{figure*}
\vspace{10.7cm}
\includegraphics{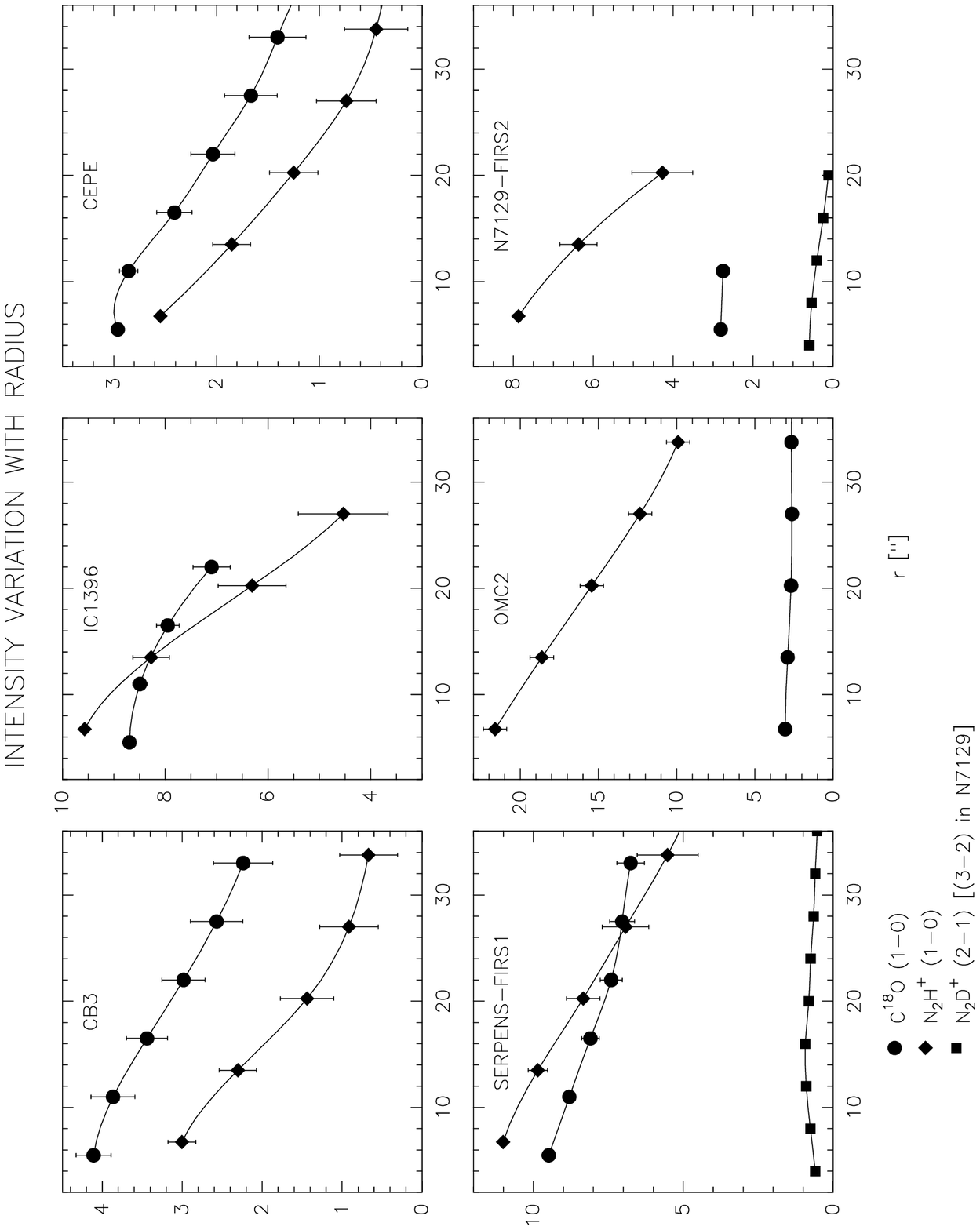}
\caption{Radially integrated intensity profiles of the C$^{18}$O 1$\rightarrow$0, N$_2$H$^+$ 1$\rightarrow$0, and N$_2$D$^+$ 2$\rightarrow$1 (3$\rightarrow$2 for NGC~7129--FIRS~2) lines in our sample of Class 0 IMs. Each point represents the mean value of the line integrated emission in a ring of a given radius. The errors are the rms of the line integrated emission in the ring. The step between two consecutive points, which is equal to the size of each ring, is equivalent to 1/4 of the beam at each frequency.
}
\end{figure*}

\section{LTE column densities}

We have derived the C$^{18}$O column densities using the rotation diagram method.
This method gives an accurate estimate of the column density provided that the emission
is optically thin, is thermalized, and arises from a homogeneous and isothermal slab.
In the case of a density and temperature distribution, the derived column density represents an average value 
over the observational beam. 

The excitation of C$^{18}$O and C$^{17}$O are very similar, and we have better signal-to-noise spectra for
the more intense  C$^{18}$O lines. For this reason, we derived the C$^{17}$O column density by assuming
optically thin emission and the rotation temperature derived from the C$^{18}$O data.
The molecule N$_2$H$^ +$ presents hyperfine splitting. This allows
us to estimate the line opacity directly from the hyperfine line ratios. 
We derived the total N$_2$H$^+$ column density  
from the opacity of the N$_2$H$^+$ 1$\rightarrow$0 line, and
assumed the rotation temperatures derived from N$_2$D$^+$
data when there was no N$_2$H$^+$ 4$\rightarrow$3 observations. The same rotation temperature was always used for both N$_2$H$^+$ and N$_2$D$^+$.
We assumed a beam filling factor of 1 for all the lines regardless of the observational beam size. 
This approach overestimates the rotation temperature when the emission is centrally peaked.
In Table 3 we present the derived C$^{18}$O, C$^{17}$O,
N$_2$H$^+$, and N$_2$D$^+$ column densities. We also show
the N(C$^{18}$O)/N(C$^{17}$O) (hereafter R$_1$) and the N(N$_2$D$^+$)/N(N$_2$H$^+$) (hereafter R$_2$) ratios.
For the whole sample, we obtain a value for R$_1$ around 3.3+/-0.3, the expected value for optically thin emission. 
In the worst case, R$_1$=2.5, the correction to the C$^{18}$O column density because of the opacity is only a factor $\sim$1.3. 
Thus, opacity effects are not important in our C$^{18}$O column density estimates.

\begin{table*}[t]
\caption{LTE column densities}
\begin{tabular}{lcccc|lccc}
\\ \hline \hline
\multicolumn{1}{l}{Source} &
\multicolumn{1}{c}{T$_{r}$} &
\multicolumn{1}{c}{N(C$^{18}$O)} &
\multicolumn{1}{c}{N(C$^{17}$O)} &
\multicolumn{1}{c|}{R$_1^  *$} &
\multicolumn{1}{c}{T$_{r}$ (K)} &
\multicolumn{1}{c}{N(N$_2$H$^+$)} &
\multicolumn{1}{c}{N(N$_2$D$^+$)} &
\multicolumn{1}{c}{R$_2^  *$} \\
\multicolumn{1}{l}{} &
\multicolumn{1}{c}{(K)} &
\multicolumn{1}{c}{(cm$^{-2}$)} &
\multicolumn{1}{c}{(cm$^{-2}$)} &
\multicolumn{1}{c|}{} &
\multicolumn{1}{c}{(K)} &
\multicolumn{1}{c}{(cm$^{-2}$)} &
\multicolumn{1}{c}{(cm$^{-2}$)} &
\multicolumn{1}{c}{} \\
\hline
Serpens-FIRS 1    & 11 & 8.2$\times$10$^{15}$  &  3.2$\times$10$^{15}$ & 2.6       &       11$^a$ &  7.7$\times$10$^{13}$ & 3.4$\times$10$^{11}$ & 0.004 \\ 
Cep E-mm            & 15 &  2.5$\times$10$^{15}$  & 7.6$\times$10$^{14}$ & 3.3     &       9$^b$ & 1.4$\times$10$^{13}$ & 5.8$\times$10$^{11}$ & 0.04 \\ 
L1641 S3 MMS 1      & 10 &  3.2$\times$10$^{15}$  & 1.3$\times$10$^{15}$ & 2.5     &       8$^b$ & 3.6$\times$10$^{13}$ & $<$2.1$\times$10$^{11}$ & $<$0.006 \\ 
IC1396N                &  25  &   1.1$\times$10$^{16}$ & 3.7$\times$10$^{15}$ & 3.0 &      10$^b$  & 4.1$\times$10$^{13}$ & $<$4.1$\times$10$^{11}$ & $<$0.01 \\ 
CB3                      & 14  &  5.2$\times$10$^{15}$ & 1.6$\times$10$^{15}$ & 3.3 &      6$^a$ & 1.2$\times$10$^{13}$ & 1.0$\times$10$^{12}$ & 0.08 \\ 
OMC2-FIR4          &  22 & 4.8$\times$10$^{15}$ & 1.2$\times$10$^ {15}$ & 3.4 &              22$^a$ & 2.0$\times$10$^{14}$ &  8.7$\times$10$^{11}$ & 0.004 \\ 
NGC 7129--FIRS 2       &  23 & 1.9$\times$10$^{15}$  & 6.9$\times$10$^{14}$ & 2.8 &        13$^c$ &  3.8$\times$10$^{13}$ & 5.4$\times$10$^{11}$ & 0.014 \\ 
S140                     &  50$^c$ & 1.0$\times$10$^{16}$  & 2.7$\times$10$^{15}$ & 3.7 &  50$^d$ & 1.2$\times$10$^{14}$ & $<$4.5$\times$10$^{11}$ &  $<$0.004 \\ 
LkH$\alpha$ 234 $^b$  &    &                         &                          &        &  22       &  1.0$\times$10$^{13}$ & $<$2.3$\times$10$^{11}$ &  $<$0.02 \\ 
\hline \hline
\end{tabular}

\noindent
$^*$ R$_1$=N(C$^{18}$O)/N(C$^{17}$O), R$_2$=N(N$_2$D$^+$)/N(N$_2$H$^+$).

\noindent
$^a$ Rotation temperature derived from our N$_2$D$^+$ observations.

\noindent
$^b$  Rotation temperature derived from our N$_2$H$^+$ observations.

\noindent
$^c$ From Fuente et al. (2005a).

\noindent
$^d$ From Minchin et al. (1995).
\end{table*}

\subsection{N$_2$D$^+$ and N$_2$H$^+$}

The deuterium fractionation of N$_2$H$^+$ (R$_2$)
strongly depends on the CO depletion factor and the gas temperature (Caselli et al. 2002,
Ceccarelli \& Dominik 2005, Daniel et al. 2007). In our sample of IM Class 0 
protostars, we derive values of R$_2$ ranging from 0.005 to 0.014. 
These values are 3 orders of magnitude higher than 10$^{-5}$, 
the elemental value in the interstellar medium (Oliveira et al. 2003). Nevertheless, the R$_2$ values are
a factor of 10 lower than those found in prestellar clumps by Crapsi et al. (2005).
According to the values of R$_2$, we can classify our sources into two groups:
(i) highly deuterated sources that have values of R$_2$$>$0.01 
and (ii) moderately deuterated sources with R$_2$$<$0.01. Cep~E--mm, CB3, and
NGC~7129--FIRS~2 belong to the first group. Assuming that the  
N(N$_2$D$^+$)/N(N$_2$H$^+$) ratio is a good gas temperature indicator,
these protostars should be the coldest and very likely the youngest of our
sample. We have to be cautious with CB3, however. Since this source is 
the most distant (d=2500 pc), our single-pointing observations
trace a larger fraction of the envelope. The sources Serpens--FIRS~1,
IC~1396~N, OMC2--FIR~4, and S140 belong to the second group. S140 is a Class I
YSO and IC~1396~N is considered a  borderline Class 0/I YSO. 
The results are thus consistent with our interpretation of the objects in this group as being more evolved.
In Serpens--FIRS~1, however, the N$_2$D$^+$ 1$\rightarrow$0 emission comes mainly from the
lower density envelope and the ratio cannot be considered a tracer of the evolutionary stage
of the protostar. OMC2~FIR 4 is also difficult to classify. The deuterium fractionation suggests an
evolved object but the morphology of the C$^{18}$O map is more consistent with a
young Class 0 star. The peculiarity of this protostellar envelope has
already been pointed out by Crimier et al. (2009). They find that the envelope of OMC2~FIR 4 is peculiarly flat
and warm with a radial density power-law index of 0.6. 

We present the N$_2$H$^+$ deuterium fractionation as a function of the N(C$^{18}$O)/N(N$_2$H$^+$)
ratio in Fig.\ 7. An excellent correlation between these two quantities is found in prestellar cores with the
deuterium fractionation decreasing with increasing N(C$^{18}$O)/N(N$_2$H$^+$)
ratio (see Crapsi et al. 2005). This correlation is not valid for IM Class  0 protostars. 
As discussed in the following sections, this is due to the complexity of these intermediate mass star-forming regions.
In prestellar cores the chemistry of these species is only driven by the CO depletion. In IMs, other phenomena like
photodissociation and shocks are also playing important roles.

\begin{table*}
\caption{Temperature-density (T-n) profiles from Crimier et al. 2010} 
\begin{tabular}{llllll} \hline
 Source                                                   & CB3 & Cep-
E\ & IC\ 1396\ N\ & NGC\ 7129\ & Serpens  \\
                                                          & mm\  & 
mm\     & BIMA2\  &  FIRS\ 2\  & FIRS\ 1          \\  \hline
\multicolumn{1}{l}{RA(J2000) } & 
\multicolumn{1}{c}{00:28:42.1} & 
\multicolumn{1}{c}{23:03:12.7} & 
\multicolumn{1}{c}{21:40:41.8} & 
\multicolumn{1}{c}{21:43:01.5} &  
\multicolumn{1}{c}{18:29:49.8} \\
\multicolumn{1}{l}{Dec(J2000)} & 
\multicolumn{1}{c}{56:41:59.4} & 
\multicolumn{1}{c}{61:42:27.4} & 
\multicolumn{1}{c}{58:16:13.5} & 
\multicolumn{1}{c}{66:03:25.0} &  
\multicolumn{1}{c}{01:15:18.4} \\ 
 Dust optical depth at 100\,$\mu$m, $\tau_{100}$          & 5.8       & 
5.0 & 1.4 & 2.3 & 3.0    \\
 Density power law index, $\alpha$                        & 2.2       & 
1.9 & 1.2 & 1.4 & 1.5    \\
 Envelope thickness, $r_{\mathrm{out}}$/$r_{\mathrm{i}}$  & 400       & 
500 & 630 & 180 & 200    \\
 \hline
 Inner envelope radius, $r_{\mathrm{in}}$, ($\mathrm{AU}$)            & 
260    &  70   & 50    & 100   & 30   \\
 Outer envelope radius, $r_{\mathrm{out}}$, ($\mathrm{AU}$)           & 
103000 & 35800 & 29600 & 18600 & 5900  \\
 Radius at T$_{dust}$ = 100 K,   $r_{\mathrm{100K}}$, ($\mathrm{AU}$) & 
700             & 223   & 180             & 373             & 102   \\
 H$_2$ density at $r_{\mathrm{100K}}$, $n_0$, ($\mathrm{cm}^{-3}$)    & 
7.5$\times$10$^{7}$ & 2.0$\times$10$^{8}$ & 4.3$\times$10$^{7}$ & 4.4$\times$10$^{7}$ & 
2.2$\times$10$^{8}$   \\
 Envelope mass, $M_{\mathrm{env}}$, ($\mathrm{M}_{\sun}$)           & 
120    &  35   & 90    & 50    & 5.0   \\
 \hline
\end{tabular}
\end{table*}

\begin{table*}[t]
\caption{Abundance profiles derived from the observations} 
\begin{tabular}{lllllllc} \\ \hline 
\multicolumn{1}{l}{\bf Serpens-FIRS 1}  &   
\multicolumn{1}{c}{X$_0$} &
\multicolumn{1}{c}{R$_0$ (AU)} &
\multicolumn{1}{c}{X$_1$} &
\multicolumn{1}{c}{R$_1$} (AU) &
\multicolumn{1}{c}{X$_2$} &
\multicolumn{1}{c}{RMS (K km s$^{-1}$)} &  \multicolumn{1}{c}{Diameter(")/HPBW(")}\\ \hline
C$^{18}$O        &   1.4$\times$10$^{-7}$   &  \multicolumn{1}{l|}{2000}  &  $<$7$\times$10$^{-9}$   &  \multicolumn{1}{l|}{6000}  &  1.1$\times$10$^{-7}$   & 0.27  & 2.3 \\
N$_2$H$^+$       &   4.0$\times$10$^{-10}$  &  \multicolumn{1}{l|}{3000}  &  $<$3$\times$10$^{-11}$  &  \multicolumn{1}{l|}{6000}  &  2.5$\times$10$^{-10}$  & 0.29  & 1.9\\  \hline 
N$_2$D$^+$       &   \multicolumn{3}{c}{$<$2$\times$10$^{-12}$}          &   \multicolumn{1}{l|}{6000}  &  1.9$\times$10$^{-11}$  & 0.6  & 3.1 \\  \hline
T$_{ev}^*$ (K)        &   \multicolumn{6}{l}{20 } \\
n$_{ev}^*$ (cm$^{-3}$) &   \multicolumn{6}{l}{2.4$\times$10$^6$} \\  \hline \hline 
%
\multicolumn{1}{l}{\bf Cep E-mm}  &   
\multicolumn{1}{c}{X$_0$} &
\multicolumn{1}{c}{R$_0$ (AU)} &
\multicolumn{1}{c}{X$_1$} &
\multicolumn{1}{c}{R$_1$} (AU) &
\multicolumn{1}{c}{X$_2$} &
\multicolumn{1}{c}{RMS (K km s$^{-1}$)} \\ \hline
C$^{18}$O  &  6.0$\times$10$^{-8}$   &   \multicolumn{1}{l|}{3500}     &  \multicolumn{3}{c}{1.1$\times$10$^{-7}$$\times$(r/35800)}    & 0.24   & 4.4 \\ 
N$_2$H$^+$ &  2.4$\times$10$^{-10}$  &   \multicolumn{1}{l|}{6000}      &  \multicolumn{3}{c}{2.5$\times$10$^{-10}$$\times$(r/35800)}    & 0.28 & 3.7 \\  \hline 
N$_2$D$^+$ &  \multicolumn{5}{c}{7.0$\times$10$^{-12}$}   &   & 6.1\\
\hline
T$_{ev}$ (K)        &   \multicolumn{6}{l}{20 } \\
n$_{ev}$ (cm$^{-3}$) &   \multicolumn{6}{l}{1.0$\times$10$^6$} \\  \hline \hline
%
%
\multicolumn{1}{l}{\bf IC 1396 N}  &   
\multicolumn{1}{c}{X$_0$} &
\multicolumn{1}{c}{R$_0$ (AU)} &
\multicolumn{1}{c}{X$_1$} &
\multicolumn{1}{c}{R$_1$} (AU) &
\multicolumn{1}{c}{X$_2$} &
\multicolumn{1}{c}{RMS (K km $s^{-1}$)} \\ \hline
C$^{18}$O   &  6.0$\times$10$^{-8}$  &  \multicolumn{1}{l|}{5500}      &  \multicolumn{3}{c}{1.25$\times$10$^{-7}$$\times$(r/29600)}   & 0.24    & 3.5 \\  \hline
N$_2$H$^+$  &  4.2$\times$10$^{-10}$ &  \multicolumn{1}{l|}{10000}    &    1.0$\times$10$^{-11}$     &    \multicolumn{1}{l|}{15000} &   1.8$\times$10$^{-10}$    & 0.50 & 3.0  \\  \hline
N$_2$D$^+$   &  \multicolumn{5}{c}{$<$7$\times$10$^{-12}$} &  & 4.9 \\ \hline
T$_{ev}$ (K)        &   \multicolumn{6}{l}{19 } \\
n$_{ev}$ (cm$^{-3}$) &   \multicolumn{6}{l}{7.0$\times$10$^5$} \\  \hline \hline
%
%
\multicolumn{1}{l}{\bf CB3}  &   
\multicolumn{1}{c}{X$_0$} &
\multicolumn{1}{c}{R$_0$ (AU)} &
\multicolumn{1}{c}{X$_1$} &
\multicolumn{1}{c}{R$_1$} (AU) &
\multicolumn{1}{c}{X$_2$} &
\multicolumn{1}{c}{RMS (K km s$^{-1}$)} \\ \hline
C$^{18}$O   & 1.3$\times$10$^{-7}$       & \multicolumn{1}{l|}{25000}        &   $<$3.0$\times$10$^{-8}$   &  \multicolumn{1}{l|}{60000}    &  9.0$\times$10$^{-7}$    & 0.24 & 3.7 \\  \hline
N$_2$H$^+$  & \multicolumn{5}{c}{6.5$\times$10$^{-10}$}    & 0.45  &  3.1 \\
N$_2$D$^+$ &  \multicolumn{5}{c}{3.0$\times$10$^{-11}$}   &  & 5.1 \\  \hline
T$_{ev}$ (K)        &   \multicolumn{6}{l}{15 } \\
n$_{ev}$ (cm$^{-3}$) &   \multicolumn{6}{l}{3.0$\times$10$^4$} \\  \hline \hline 
%
%
\multicolumn{1}{l}{\bf NGC 7129--FIRS 2}  &   
\multicolumn{1}{c}{X$_0$} &
\multicolumn{1}{c}{R$_0$ (AU)} &
\multicolumn{1}{c}{X$_1$} &
\multicolumn{1}{c}{R$_1$} (AU) &
\multicolumn{1}{c}{X$_2$} &
\multicolumn{1}{c}{RMS (K km s$^{-1}$)} \\ \hline
C$^{18}$O   &  \multicolumn{5}{c}{4.0$\times$10$^{-8}$} & 0.4 & 1.4  \\  
N$_2$H$^+$  &  \multicolumn{5}{c}{4.7$\times$10$^{-10}$} & 0.4 & 1.1  \\ \hline                 
N$_2$D$^+$  &  $<$3.0$\times$10$^{-12}$ & \multicolumn{1}{l|}{9000} &    4.5$\times$10$^{-11}$   &  \multicolumn{1}{l|}{11000}  & $<$3.0$\times$10$^{-12}$        & 0.13 & 2.7  \\ \hline
\hline
\end{tabular}

\noindent 
$^*$ Temperature and density at the CO evaporation radius.
\end{table*}

\section{Radiative transfer model}

We utilized a general ray-tracing radiative transfer code\footnote{The model is called DataCube and a link to install it is available upon request.}
to derive the fractional abundance profiles of C$^{18}$O, N$_2$H$^+$ and N$_2$D$^+$ across the envelopes. Assuming appropriate radial profiles for the temperature, density, molecular abundance, and turbulence velocity, this  model calculated the brightness temperature distribution on the sky.
The model map was then convolved with the telescope beam profile.
The underlying source geometry was assumed to be a sphere, with the inner and outer radii, and temperature-density (T-n) profiles derived by C10. The size of the grid was set to 32x32 cells.
The cells have different sizes along the line of sight to account for the different slopes in the density and temperature profiles. We used very small cells ($\lesssim$100 AU) near the center, where the temperature and density gradients are highest. In the outer regions, the cells reach several thousand AU in size. 
The turbulent velocity was assumed to be fixed at 1.5 km~s$^{-1}$, consistent with the linewidths of the observed lines. Finally,
in each cell, the excitation temperature was calculated with the RADEX code (van der Tak et al. 2007), which uses the slab LVG approximation at each shell.  We used the collisional rates provided by the LAMDA database\footnote{LAMDA database is available at http://www.strw.leidenuniv.nl/$\sim$moldata/.} (Sch\"oier et al. 2005). For N$_2$D$^+$, we used the same collisional rates as for N$_2$H$^+$    .

With these assumptions, we searched for the best fit on every source and molecule observed. The fitting process consisted of averaging the observed and modeled fluxes in concentric rings around the center position using GILDAS software. The center position was selected to be the continuum emission peak
at 850~$\mu$m. 
The radius of the first ring was set to HPBW/4, and incremented by the same amount in consecutive rings.  As a first step we searched for the best fit using a constant abundance. Since this approach seldom produced good fits, we next searched for the best fit using radial power functions and step functions. These functions were not selected arbitrarily but were selected to mimic the predictions of chemical models. The step function accounts for the abrupt sublimation of the CO ices thanks to the increase in the dust temperature going inwards, whereas the power-law profile accounts for a smoother change in the abundance of the species. The angular resolution of our observations ($\sim$16$"$--27$"$ depending on the frequency, i.e, $\sim$16000~AU--27000~AU at a distance of 1000~pc) prevents us from tracing the inner region (R$<\ $a few 1000~AU) of the protostellar envelope.
For this reason, we assume a constant abundance in the inner part. 

We fit the integrated intensity maps of the C$^{18}$O 1$\rightarrow$0, N$_2$H$^+$ 1$\rightarrow$0, N$_2$H$^+$ 4$\rightarrow$3, and N$_2$D$^+$ 2$\rightarrow$1 lines, when possible. The maps of the C$^{18}$O 2$\rightarrow$1 and C$^{17}$O 1$\rightarrow$0 and 2$\rightarrow$1 lines are of less quality so we did not use them in our fits.
Some sources show deviations from the spherical symmetry because of the contribution of the surrounding molecular cloud (Serpens--FIRS~1, CB3) or because of the cavity excavated by the outflow (IC~1396~N). In these cases we masked part of the map to avoid this contamination in the rings. A short description of earch source and the individual details about the modeling are given in Appendix A. The areas masked are shown in Fig. 4, and the abundance profiles obtained from the model are listed in Table~5.
We did not model the Class I sources S140 and LkH$\alpha$~234. The modeling of OMC2~FIR~4 will be the subject of a forthcoming paper. 

\section{Limitations of our model: CO evaporation temperature }

The model used here is not a reliable predictor of the chemical and physical properties within
the inner envelope for several reasons. First of all we only considered the low-J rotational molecular lines.
The emission of these transitions arises mainly from the outer envelope. This fact, together with the limited 
spatial resolution of our single dish observations, prevent us from determining the variation in the abundance
in the inner envelope. Additionally, in our modeling we use the n-T profiles derived by C10. These profiles are a reasonable
 approximation for the outer envelope but are relatively unconstrained in the inner region. 
Some parameters, like the inner radius of the envelope and the dust temperature at this radius,
are thus poorly determined. 

Another important source of uncertainty is the degeneracy of the solutions. For instance, in the case of
C$^{18}$O we can obtain similar results by varying the molecular abundance in the inner region (X$_0$) or the CO evaporation radius (R$_0$), as long as the line is optically thin and the number of molecules in the beam remains constant. 
Here we discuss the impact of this degeneracy in the derived CO evaporation radius.

We ran a grid of models for Cep~E-mm, IC~1396~N, and
CB3. These are the sources in which the ratio between the angular diameter of the envelope and the HPBW of 
the telescope is greatest, so we are more sensitive to the spatial variations of the chemical abundances. 
In all these models, we assume that the C$^{18}$O abundance profile is X=X$_0$ for radii less than 
the evaporation radius, R$_0$,  and X= X$_0$$\times$(R/R$_{out}$)$^\alpha$  for radii larger than R$_0$. For each value
of X$_0$, we fit R$_0$ and $\alpha$. X$_0$ is the un-depleted C$^{18}$O abundance, R$_0$ the evaporation radius, and $\alpha$
measures the gradient in the C$^{18}$O abundance due to depletion. 
The canonical abundance of C$^{18}$O in principle depends on the Galactocentric distance (DGC). Following
Wilson \& Matteucci (1992), the CO abundance is given by
\begin{equation}
X_{(CO)} = 9.5\times 10^{−5} exp (1.105 −- (0.13 \times DGC(kpc))).
\end{equation}

Following Wilson \& Rood (1994), the oxygen isotope ratio $^{16}$O/$^{18}$O depends on DGC according to the relationship
$^{16}$O/$^{18}$O~=~58.8~$\times$~DGC\,(kpc)~+~37.1. At the distance of the Sun, 8.5~kpc,
X$_{C^{18}O}$=1.7$\times$10$^{-7}$. This is an average value, so there may be local effects.
In our models we varied X$_0$ from 1$\times$10$^{-7}$ to 4$\times$10$^{-7}$,
the range of X$_0$ values that we consider reasonable. For each value of X$_0$, we varied R$_0$ and $\alpha$ to find the best fit.
In Fig. 8 we show the results of our models for Cep~E. The best fit is
X$_0$ = 1$\times$10$^{-7}$, R$_0$=2510~AU, $\alpha$=1 with RMS=0.37~K~km~s$^{-1}$.
We consider that the models with RMS values departing by a factor of less than 1.3 from the minimum value are still acceptable. Following this criterion,
we have two different families of solutions. Assuming X$_0$=1$\times$10$^{-7}$,
the best fit is obtained with $\alpha$$\sim$1 and R$_0$$\sim$2100--3100~AU. Assuming  X$_0$=2$\times$10$^{-7}$,
we have reasonably good solutions with  $\alpha$$\sim$2 and R$_0$$\sim$2200--3200~AU. We have not
found any acceptable solution for higher values of X$_0$. The evaporation radius of C$^{18}$O is thus 2600$\pm$600~AU.
These radii correspond to dust temperatures of 20-25~K.
Interestingly, we would have a better fit to the observations (RMS=0.24~K~km~s$^{-1}$) if we allowed the
C$^{18}$O abundance in the inner region to fall below 1$\times$10$^{-7}$. This is the solution we show
in Table 5. We do not reject this solution because 
there are several physical reasons that the CO abundance could be lower in the inner envelope. First of
all, C$^{18}$O could have been photodissociated in the regions very close to the recently formed star.
Alternatively, part of the CO could have been transformed into CH$_3$OH on the icy mantles, and the CO abundance once
evaporated would then be different from the initial value. However, given the large uncertainty in the physical structure of the
inner parts of the envelope, we cannot draw any conclusion. 

In Fig. 9 we show the results for IC~1396~N. The best fit is 
X$_0$~=~1$\times$10$^{-7}$, R$_0$=2000~AU, and $\alpha$=0.5 with  RMS=0.36~K~km~s$^{-1}$. 
We consider that the models with values of RMS$<$0.5 are acceptable. Following this criterion,
we only find solutions for X$_0$=1$\times$10$^{-7}$ and R$_0$$<$3000~AU; i.e., the CO evaporation
temperature should be $>$23~K. Again, the fit can be significantly improved
if values of X$_0$ lower than 1$\times$10$^{-7}$ are considered. This is similar to
the case of Cep E. Finally, in Fig. 10, we show the same grid of models for CB3.
The best fit is for X$_0$ = 4$\times$10$^{-7}$, R$_0$=4000~AU, and $\alpha$=0.76 with  
RMS=2.0~K~km~s$^{-1}$, which corresponds to a CO evaporation temperature of $\sim$30~K.
However, this is not a good fit. In fact, we obtain a much better solution when assuming
a step function and T$_{ev}$$\sim$15~K (see Table 5). 

For Serpens-FIRS~1 and NGC~7129~FIRS2, the protostars are barely resolved by our single-dish observations. In addition,
the contribution of the surrounding cloud is very significant. For these reasons, we did not carry out a study similar to what is described above. We
looked only for abundance profiles that fit our observations and are consistent with the behavior predicted by chemical models (see Table~5).
In these two cases, we can fit the observations when assuming that the CO is evaporated at temperatures around 20--25~K. We conclude, therefore, that
our C$^{18}$O observations towards the Class 0 stars are better fit assuming CO evaporation temperatures of 20--25~K that are consistent
with the CO being bound in a CO-CO matrix.

The fits to the N$_2$H$^+$ and N$_2$D$^+$ emission profiles have been done by hand. For N$_2$H$^+$ we tried two options:(i) constant 
N$_2$H$^+$ abundance and (ii) an abundance profile with a radial variation similar to that of C$^{18}$O. Option (ii) follows the expectation that the N$_2$H$^+$ abundance is strongly dependent on the CO abundance. In the case of CB3 and NGC~7129~FIRS~2, we are able to fit the N$_2$H$^+$ observations by assuming a constant
abundance. For the rest of sources, the N$_2$H$^+$ observations are better fit by assuming that the N$_2$H$^+$ is depleted in the cold regions similarly
to CO.

One expects an annular distribution for the N$_2$D$^+$ abundance, with the maximum N$_2$D$^+$ abundance in the region with the greatest CO depletion.
The morphology observed in the integrated intensity map of the  N$_2$D$^+$ 2$\rightarrow$1 line towards Serpens-FIRS 1 suggests that the
N$_2$D$^+$ emission is dominated by the foreground molecular cloud. For this reason we selected a one-step function with a constant
abundance inside and outside the protostellar core. In the case of NGC~7129--FIRS 2, we detected a clump of N$_2$D$^+$ 3$\rightarrow$2 emission towards 
the star. However, interferometric observations show that on scales of a few 1000~AU, there is a lack of emission towards the star (Fuente et al 2005a). We thus fit the N$_2$D$^+$ emission assuming an annular abundance distribution.

\setlength\unitlength{1cm}
\begin{figure}
\vspace{6cm}
\includegraphics{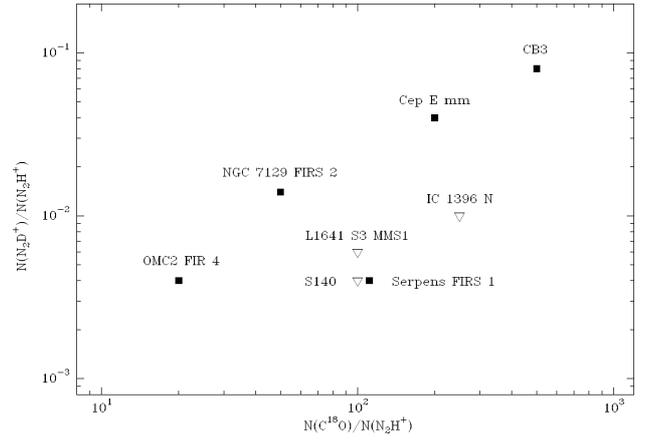}
\caption{Deuterium fractionation of N$_2$H$^+$(R$_2$) vs the N(C$^{18}$O)/N(N$_2$H$^+$) ratio in the studied YSOs. Inverted empty triangles are upper limits for L1641~S3~MMS1, IC~1396~N, and S140, where we did not detect the N$_2$D$^+$ 2$\rightarrow$1 line.}
\end{figure}

\setlength\unitlength{1cm}
\begin{figure}
\vspace{7.5cm}
\includegraphics{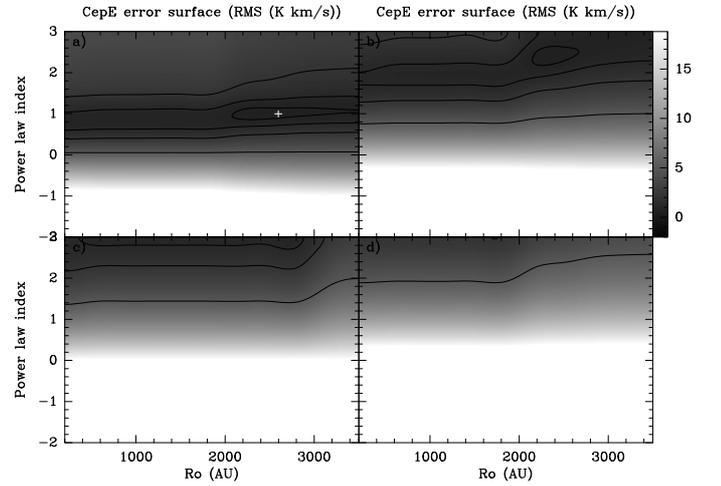}
\caption{Plots of the RMS, defined as $\Sigma (I_{model}-I_{obs})^2$) for the grid of models
run to reproduce the C$^{18}$O emission in Cep E-mm. The C$^{18}$O abundance is assumed to be
X=X$_0$ for radii less than R$_0$,  and X=~X$_0$$\times$(R/R$_{out}$)$^\alpha$  for radii larger 
than R$_0$. The values of X$_0$ are 1.0$\times$10$^{-7}$ (panel a.), 2.0$\times$10$^{-7}$ (panel b.),
3.0$\times$10$^{-7}$ (panel c.), and 4.0$\times$10$^{-7}$ (panel d.). The solution with the lowest RMS,
0.37, is indicated by a cross. We consider acceptable solutions those with RMS less than 0.5. Contour 
levels are 0.5, 1, 2, and 5 K~km~s$^{-1}$.}
\end{figure}

\setlength\unitlength{1cm}
\begin{figure}
\vspace{7.5cm}
\includegraphics{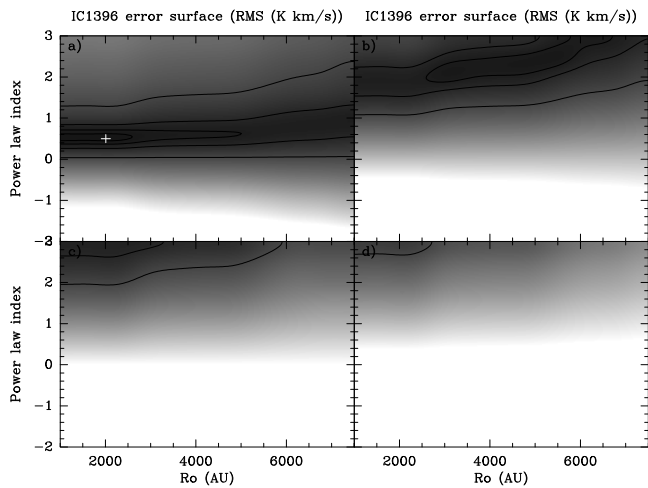}
\caption{The same as Fig. 8 for IC~1396~N. The solution with the lowest RMS,
0.36, is indicated by a cross. We consider acceptable solutions those with RMS less than 0.5. Contour 
levels are 0.5, 1, 2, and 5 K~km~s$^{-1}$.}
\end{figure}

\setlength\unitlength{1cm}
\begin{figure}
\vspace{7.5cm}
\includegraphics{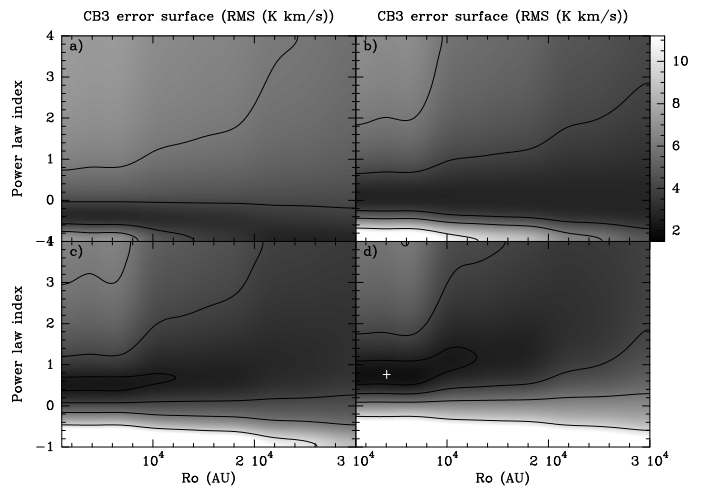}
\caption{The same as Fig. 8 for CB~3. The solution with the lowest RMS,
2.5, is indicated by a cross.  Contour 
levels are 2.5, 4, 6, and 10 K~km~s$^{-1}$.}
\end{figure}

\begin{table*}[t]
\caption{Model results}
\begin{tabular}{llllccc} \\  \hline \hline
\multicolumn{1}{c}{Source} &
\multicolumn{1}{c}{Model} &       
\multicolumn{1}{c}{CO Binding energy} &       
\multicolumn{1}{c}{$\chi$$_{C^{18}O}$}$^1$ &
\multicolumn{1}{c}{$\chi$$_{N_2H^+}$} &
\multicolumn{1}{c}{$\chi$$_{N_2D^+}$} &
\multicolumn{1}{c}{} 
 \\ \hline
Serpens--FIRS 1   &  Model 3 & 1100   &  4.37  &  3.89  &  4.52  &  only 10\% of CO evaporated \\
                  & {\bf Model 4} & {\bf 1100}   &  {\bf 2.55}  &  {\bf 3.77}  &  {\bf 4.46}  &  {\bf only 10\% survives for T$>$100~K} \\ \\
Cep E-mm    &  Model 1 & 1100   & 1.48  & 4.12  & 0.71  &  \\ 
            &  Model 2 & 5000   & 1.51  & 0.86  & 0.45  &  \\ 
            &  {\bf Model 3} & {\bf 1100}   & {\bf 0.91}  & {\bf 2.26}  & {\bf 1.19}  &  {\bf only 10\% of CO evaporated} \\
            &  Model 4 & 1100   & 1.42  & 2.19  & 1.12  &  only 10\% survives for T$>$100~K \\ \\
IC~1396~N   &  Model 3 & 1100   & 2.95  & 4.45  & $>$4.3    &  only 10\% of CO evaporated \\
            &  {\bf Model 4} & {\bf 1100}   & {\bf 1.79}  & {\bf 4.35}  & {\bf $>$4.0}    &   {\bf only 10\% survives for T$>$100~K} \\ \\
CB3         &  {\bf Model 3} & {\bf 1100}   & {\bf 1.63}  & {\bf 0.76}  & {\bf 0.31}  &  {\bf only 10\% of CO evaporated} \\
            &  Model 4 & 1100   & 1.24  & 0.86  & 0.51  &  only 10\% survives for T$>$100~K \\ \\
NGC~7129--FIRS~2   &  {\bf Model 3} & {\bf 1100}   &  {\bf 0.97} &  {\bf 1.29}  &  {\bf 2.05}   &  {\bf only 10\% of CO evaporated} \\
                   &  Model 4 & 1100   &  1.39 &  1.48  &  1.73   &  only 10\% survives for T$>$100~K \\ \\
\hline \hline
\end{tabular}

\noindent
$^1$ $\chi$ is defined as $\chi$=$\sqrt[]{[\Sigma (I_{model}-I_{obs})^2]/N} $. Note that this is an absolute error and the comparison
among values in different sources is not\\
straightforward. See Figs. 11 to  18.
\end{table*}

\section{Chemical model}
 
The chemical composition was modeled with the simple chemical 
code originally described in Caselli et al. (2002),
and updated following Caselli et al. (2008) with new measurements of 
the CO and N$_2$ binding energies (Collings et al. 2003;
\"Oberg et al. 2005) and sticking coefficients (Bisschop et al.
2006), thermal desorption, and the detailed physical structure.
The clouds are assumed to be spherically symmetric, with the density 
and temperature profiles derived from the dust continuum emission. 
An interpolation procedure has been included in
the code in order to have smooth profiles. The chemical network contains 
the neutral species CO and N$_2$, which can freeze out
onto dust grains and desorb owing to cosmic ray impulsive heating 
(as in Hasegawa \& Herbst 1993) and by thermal evaporation 
(following Hasegawa et al. 1992). The initial abundances
of CO and N$_2$ have been fixed to 9.5$\times$10$^{-5}$ (Frerking et al.
1982) and 2$\times$10$^{-5}$, respectively. 
The abundances of molecular and atomic nitrogen are difficult to determine in dense cores, but recent works suggest low values in low-mass, star-forming regions: X(N)~=~n(N)/n(H$_2$)~$\lesssim$~2$\times$10$^{-6}$ (Hily-Blant et al. 2010), and X(N$_2$)$\sim$10$^{-6}$ (Maret et al. 2006).  Our adopted value is consistent with 25\% of the total abundance of N (n(N)/n(H) = 7.9$\times$10$^{-5}$, see Anders \& Grevesse 1989) locked in N$_2$ and with the pseudo-time dependent models of Lee et al. (1996), after the chemistry reaches steady state.
Although atomic oxygen can affect the amount of deuterium fractionation (see discussion in Caselli et al. 2002), no atomic oxygen is included in the code because of the large 
uncertainties associated with its value (see e.g. Caux et al. 1999 and
Melnick \& Bergin 2005). This issue is discussed further at the end of this section.

The abundances of the molecular ions (HCO$^+$, N$_2$H$^+$, H$_3^+$, and 
all their deuterated forms) were calculated in terms of the instantaneous abundances of 
neutral species, assuming that the timescale for ion chemistry is much shorter than 
for freeze-out (Caselli et al. 2002). The rate coefficients are adopted from the UMIST 
database (http://www.udfa.net). For the proton-deuteron exchange reactions (such as 
H$_3^+$ + HD $\rightarrow$ H$_2$D$^+$ + H$_2$), we used the rates measured by 
Gerlich et al. (2002), which better fit the deuterium fractionation in low-mass Class 0 
sources, as recently found by Emprechtinger et al. (2009). Hugo et al. (2009) have recently measured the proton-deuteron rate coefficients again, finding average values of total rates (for H$_3^+$ and its deuterated isotopologues, the total rate refers to the average of multiple rates weighted according to the fraction of ortho and para forms; Sipil\"a et al. 2010) about 4-5 times more than those derived by Gerlich et al. (2002, see also Sipil\"a et al. 2010). Although this difference could affect our results by increasing the deuterium fractionation by a similar factor (thus worsening the comparison with observations), we decided not to include the new values because the nuclear spin variants of all H$_3^+$ isotopologues and H$_2$ have not been distinguished in the current model. In fact, as Pagani et al. (1992) and Flower et al. (2004) showed, small temperature variations significantly alter the ortho-to-para ratio of H$_2$, which in turn strongly affects the D-fractionation (ortho-H$_2$ being more efficient than para-H$_2$ in driving back the proton-deuteron exchange reactions thus decreasing the D-fractionation), even in the temperature regime between 9 and 20~K, where CO is mostly frozen onto dust grains (see also the discussion about the drop in ortho-H$_2$D$^+$ column density at temperatures above 10~K in Caselli et al. 2008). Such temperature variations are definitely present in the envelopes of intermediate-mass Class 0 protostars (see also Emprechtinger et al. 2009), so that a simple increase in the rate coefficients without accounting for nuclear spin variants and, in particular, the possible increase in the H$_2$ ortho-to-para ratio may overestimate the D-fractionation calculated by our models. A more detailed chemical network is currently under development.
The electron fraction has 
been computed using a simplified version of the reaction scheme of Umebayashi \& Nakano
(1990), where a generic molecular ion mH$^+$ is formed via
proton transfer with H$_3^+$, and it is destroyed by dissociative recombination 
with electrons and recombination on grain surfaces (using rates from Draine \& Sutin 1987). 
Dust grains follow a Mathis et al. (1977; MRN) size distribution, but the minimum size 
has been increased to 5$\times$10$^{-6}$ cm to simulate possible dust coagulation, as in 
the best-fit model of the prestellar core L1544 shown by Vastel et al. 2006 (see also Flower et al.
2005 and Bergin et al. 2006). The initial abundance of metals 
(assumed to freeze out with a rate similar to that of CO) is 10$^{-6}$ (see McKee 1989). 
The cosmic ray ionization rate is fixed at $\zeta$=3$\times$10$^{-17}$~s$^{-1}$ (van der Tak \&
van Dishoeck 2000). The adopted CO binding energy is 1100~K, a weighted mean of the CO 
binding energy on icy mantles (1180~K; Collings et al. 2003) and CO mantles (885~K; \"Oberg et al. 
2005), assuming that solid water is about four times more abundant than solid CO. The models are 
computed for envelope lifetimes of 10$^6$ yr, although solutions do not appreciably change after  $\simeq$10$^5$ yr.

\setlength\unitlength{1cm}
\begin{figure*}
\vspace{5.25cm}
\includegraphics{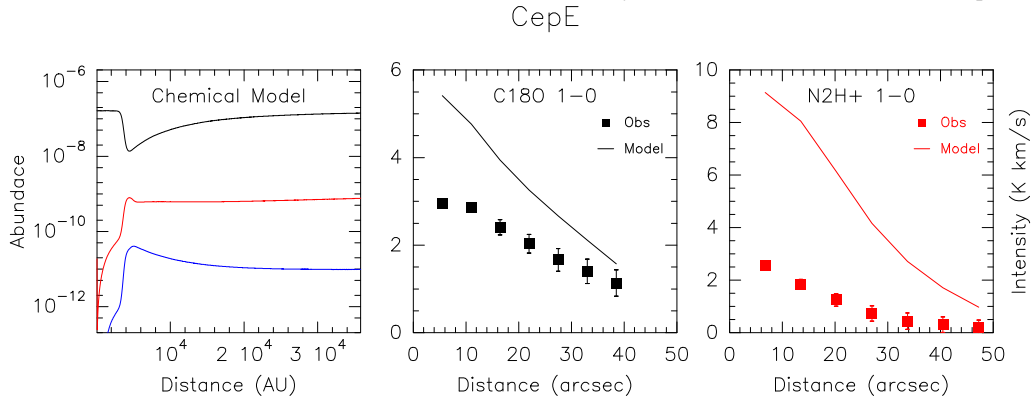}
\caption{{\it Left:} results of our chemical model for Cep~E-mm assuming a binding energy of 1100 K (standard
value, model 1). The C$^{18}$O abundance is shown in black, N$_2$H$^+$ in red, and N$_2$D$^+$ in blue. 
{\it Center:}  Comparison between the predicted C$^{18}$O 1$\rightarrow$0 line intensities (continuous line) and 
the observed values (filled squares).
{\it Right:}  Same for N$_2$H$^+$ 1$\rightarrow$0. 
}
\end{figure*}

\setlength\unitlength{1cm}
\begin{figure*}
\vspace{5.15cm}
\includegraphics{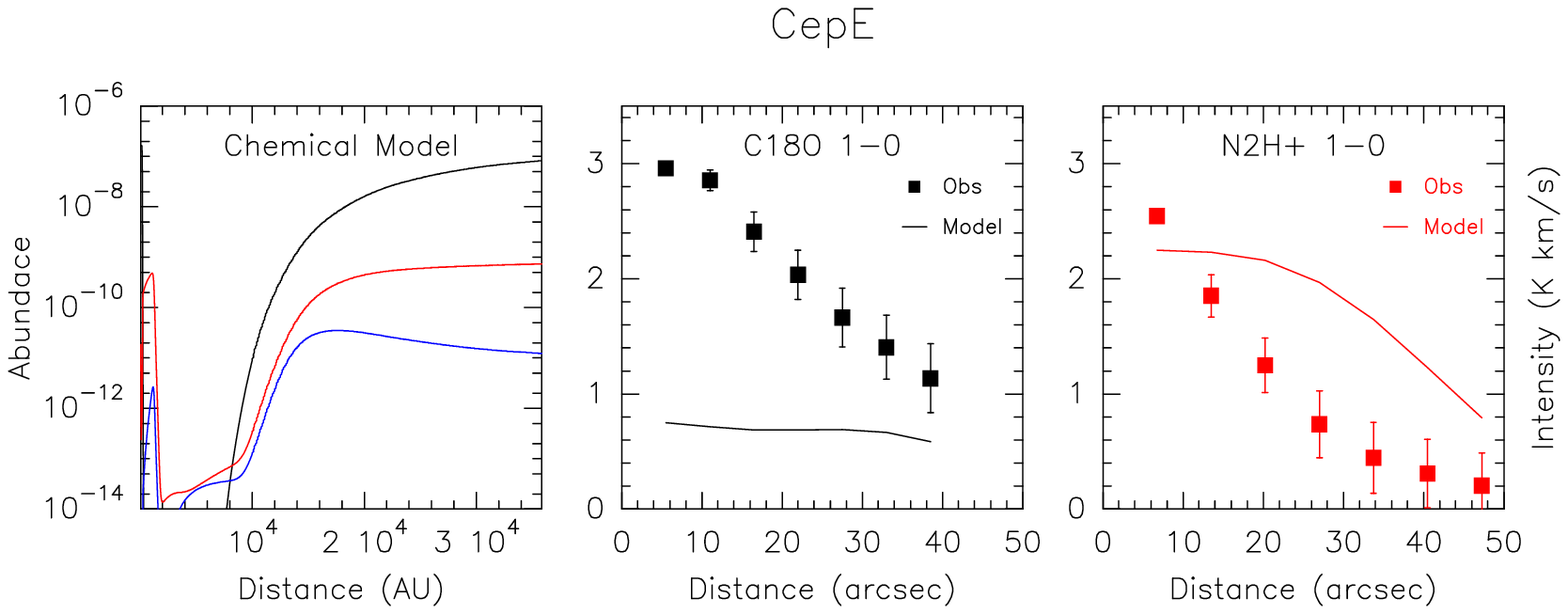}
\caption{The same as Fig. 11 for model 2.}
\end{figure*}

\setlength\unitlength{1cm}
\begin{figure*}
\vspace{5cm}
\includegraphics{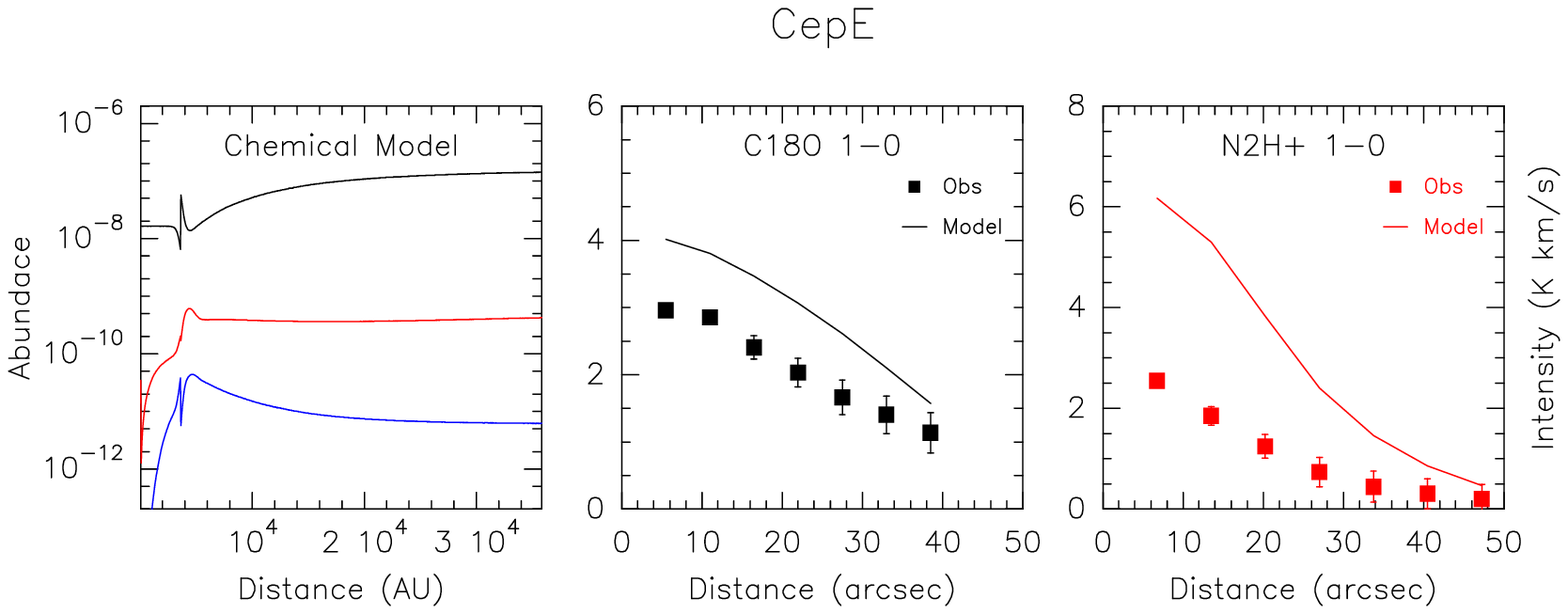}
\caption{The same as Fig. 11 for model 3.}
\end{figure*}

\setlength\unitlength{1cm}
\begin{figure*}
\vspace{5cm}
\includegraphics{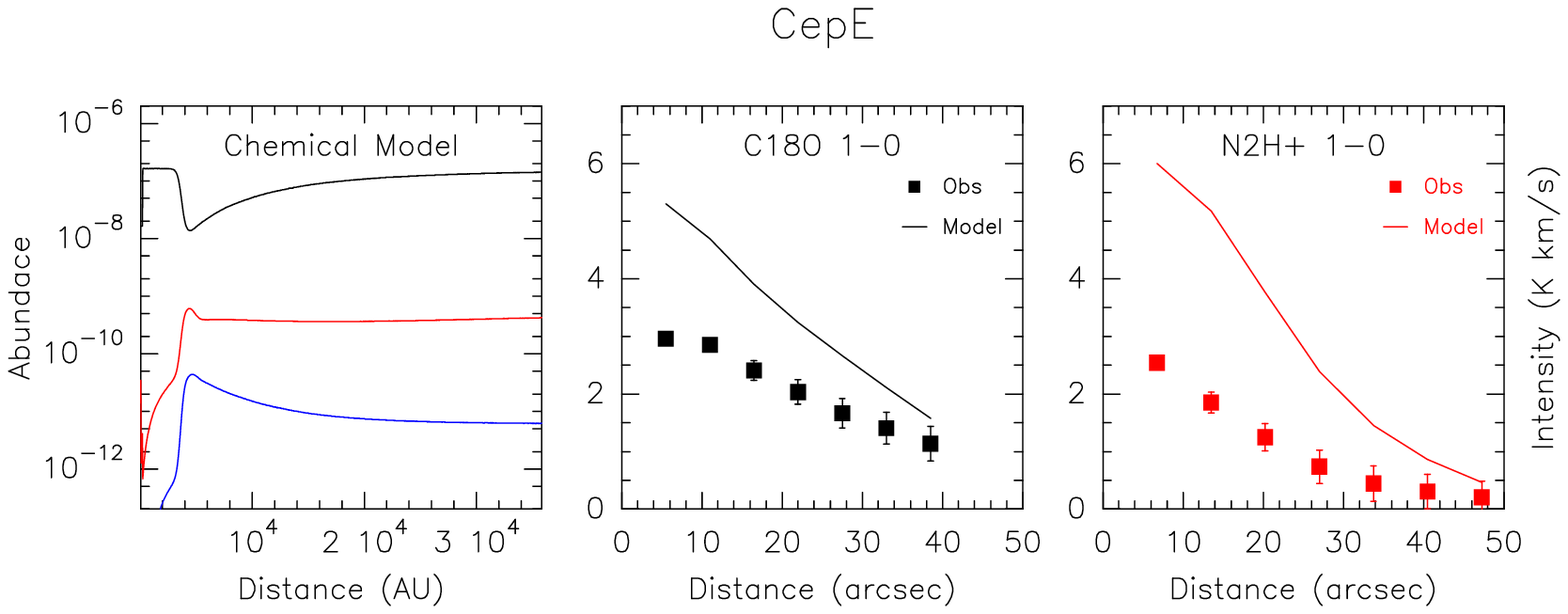}
\caption{The same as Fig. 11 for model 4.}
\end{figure*}

\setlength\unitlength{1cm}
\begin{figure*}
\vspace{5.3cm}
\includegraphics{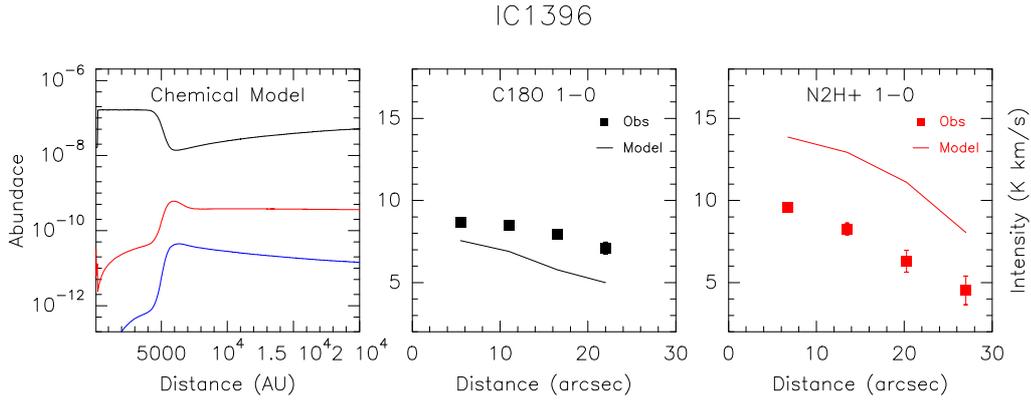}
\caption{The same as Fig. 11 for IC~1396~N and model 4.}
\end{figure*}

\setlength\unitlength{1cm}
\begin{figure*}
\vspace{5.2cm}
\includegraphics{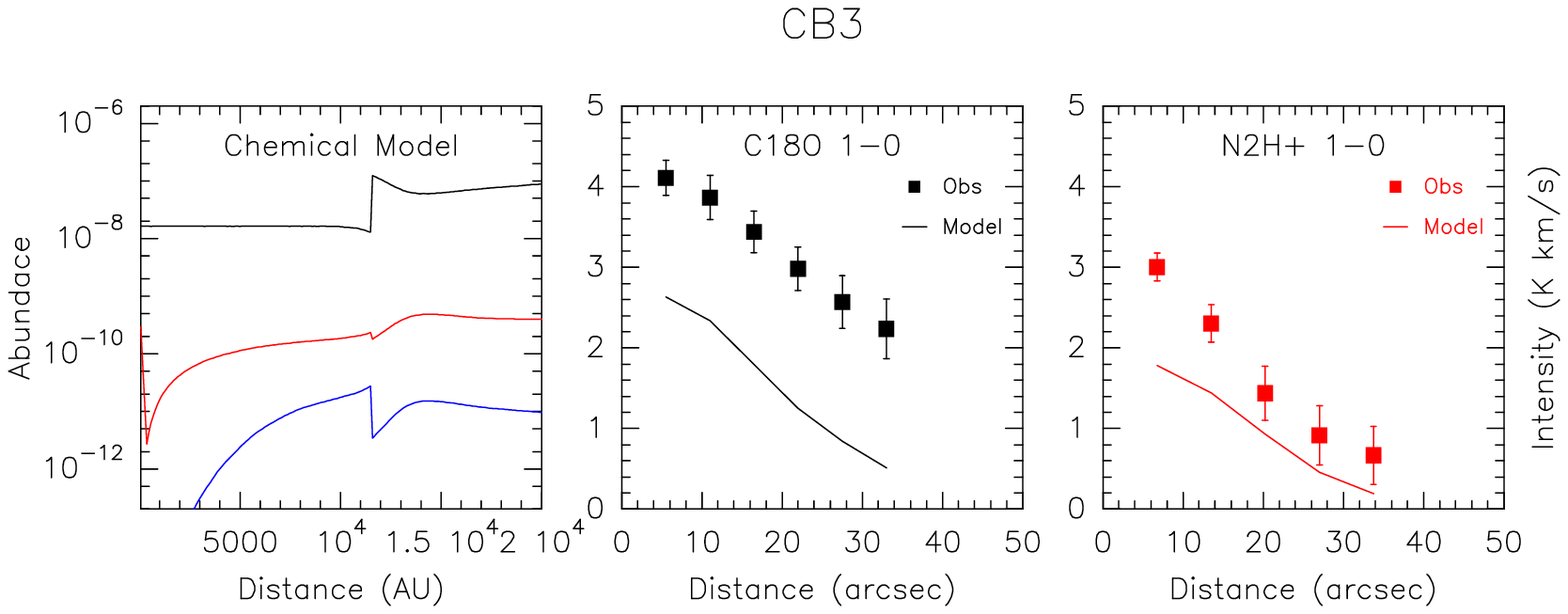}
\caption{The same as Fig. 11 for CB~3 and model 3.}
\end{figure*}

\setlength\unitlength{1cm}
\begin{figure*}
\vspace{5.2cm}
\includegraphics{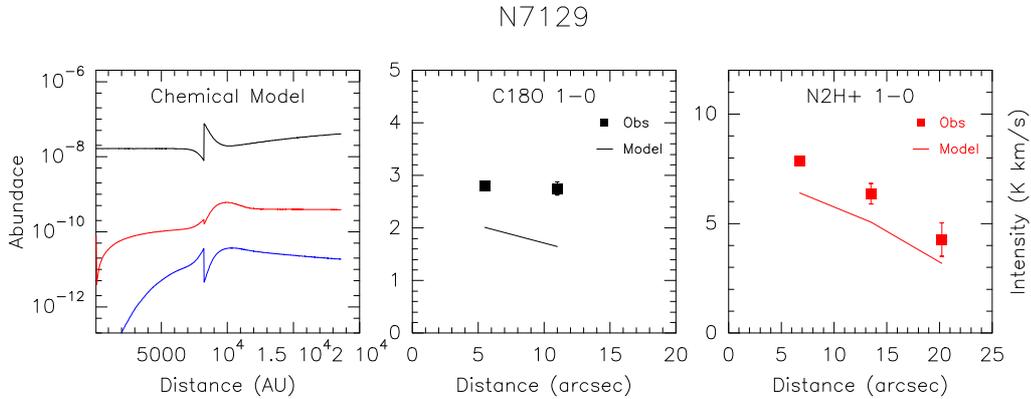}
\caption{The same as Fig. 11 for NGC~7129--FIRS~2 and model 3.}
\end{figure*}

\setlength\unitlength{1cm}
\begin{figure*}
\vspace{5.2cm}
\includegraphics{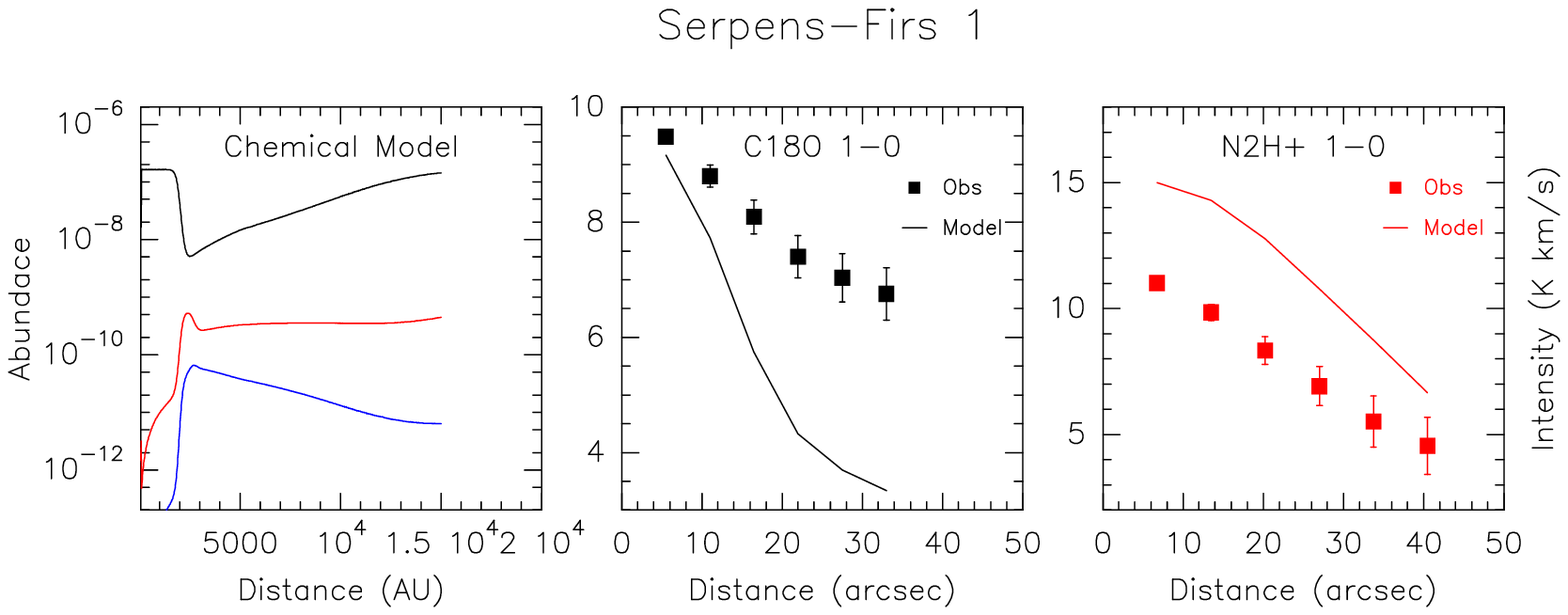}
\caption{The same as Fig. 11 for Serpens--FIRS~1 and model 3.}
\end{figure*}

In Figs. 11 to 14, we show the fits for Cep E-mm. We adopt this protostar as a fiducial example because it is the one with the best 
spatial sampling of the envelope, its geometry looks spherical, and the contribution 
of the surrounding molecular cloud is negligible.  It is impossible to reproduce the observations 
towards Cep~E-mm using the standard chemical model. With the standard model, i.e., assuming a CO binding 
energy of $\sim$1100~K, we are able to reproduce the qualitative behavior of the C$^{18}$O and 
N$_2$H$^+$ emission (see Fig. 11), but the model reproduces the line integrated 
intensities poorly. The line intensity predictions are a factor of 2--4 higher than the observations
(see Fig. 11). Increasing the CO binding energy (e.g. assuming that a large 
fraction of CO is trapped in water ice) does not improve the fit (see model 2 in Table 6 and Fig. 12). 

The abundance profiles given in Table 5 suggest that the fit to the C$^{18}$O emission would improve
by lowering the CO abundance in the inner part of the core. As pointed out there, this low CO abundance has some physical justification. One possibility is that a significant fraction of CO is converted 
into CH$_3$OH on the grain surfaces before evaporation, in agreement with observations of solid methanol 
along the line of sight towards embedded young stellar objects (e.g. Boogert et al. 2008). We mimic this situation with model 3. In this model,
only 10\% of the CO is released back to the gas phase at the CO evaporation temperature. Another possibility
is that CO is either destroyed by the stellar UV radiation and/or X-rays close to the star or transformed into more 
complex molecules via hot core chemistry.  This case corresponds to our model 4, where only 10\% of the CO survives 
when the temperature is higher than 100~K.
The two models, 3 and 4, fit the C$^{18}$O and N$_2$H$^+$ emission better than the standard model (see Figs. 13 and
14), with model 3 being a slightly better fit to the Cep E-mm observations. Thus, we applied these most successful models, model 3 and 4, to all the other sources
and have obtained reasonable fits (line integrated intensities fitted within a factor of 2) to the C$^{18}$O and N$_2$H$^+$ emission 
in all of them. For both IC~1396~N and Serpens--FIRS~1, model 4 gives a better fit, while model 3 is better for the rest.

While we obtain reasonable fits for C$^{18}$O and N$_2$H$^+$, our models do not succeed in reproducing the N$_2$D$^+$ data towards
Serpens--FIRS~1 and IC~1396~N. In fact, the chemical models predict intensities higher by a factor of $\sim$10 than the 
observed intensities for these sources. All the considered models  predict that the spatial distribution of N$_2$D$^+$ 
is similar to that of N$_2$H$^+$ and that the deuteration fraction [N$_2$D$^+$]/[N$_2$H$^+$]$\sim$0.01 in most of the envelope.
This is not true for our low-deuterated sources, Serpens--FIRS~1 and IC~1396~N, in which the  
[N$_2$D$^+$]/[N$_2$H$^+$] ratio is a few 0.001. This discrepancy could have different origins. First of all, the models 
assume a spherical geometry. It is clear that bipolar outflows have excavated large cavities in these protostellar 
envelopes (see e.g. Fuente et al. 2009). The walls of these cavities are warmed by the stellar UV radiation and shocks. 
Moderate temperatures, UV radiation, and shocks would lower the abundance of N$_2$H$^+$ and change the [N$_2$D$^+$]/[N$_2$H$^+$] ratio.
In fact, the large linewidths observed ($\sim$1.5 km s$^{-1}$) are hard to maintain without
shocks and dissipation. The role of UV radiation and shocks is expected to be greater in IM Class 0 objects than in low-mass ones.

This simple model ignores many other important parameters that could decrease the deuterium fractionation in warmer sources. These include
(i) an increased abundance of atomic O in the gas phase, possibly coming from the release of water from the icy dust mantles. Atomic oxygen is 
an efficient destruction partner for all the H$_3^+$ isotopologues, and lowers the D-fractionation, as discussed by  Caselli et al. (2002). 
(ii) The ortho-to-para H$_2$ ratio, which in systems out of equilibrium (such as free-falling) can exceed the steady-state 
value by more than an order of magnitude (Flower et al. 2006, Pagani et al. 2009). A larger fraction of ortho-H$_2$ leads to a lower D-fractionation, given that the more 
energetic ortho-H$_2$ can more easily drive the proton-deuteron exchange reactions (e.g. H$_3^+$ + HD $\rightarrow$ H$_2$D$^+$ + H$_2$) backward and reduce the H$_2$D$^+$/H$_3^+$ abundance ratio (Gerlich et al. 2002). (iii) A higher ionization rate, $\zeta$, which may be due to the presence of 
X-rays.  A larger $\zeta$ implies a larger electron fraction and thus a higher dissociative recombination rates for molecular ions, including H$_3^+$ isotopologues (e.g. Caselli et al. 2008).  (iv) The presence of small grains (in particular PAHs; see discussion in Caselli et al. 2008, Sect. 5.1) may arise from the interaction of outflow lobes and UV radiation with the molecular envelope.  The associated shocks will partially destroy dust grains along the way (e.g. Jones et al. 1996; Caselli et al. 1997; Guillet et al. 2009) and increase the surface area for recombination of molecular ions onto dust grains. 

Because of the low angular resolution of the present observations, it is hard to disentangle the influences of the above parameters. Both more detailed observations and more comprehensive models are needed for a deeper understanding of the chemical and physical evolution of the envelopes surrounding young intermediate-mass stars.

\section{Summary and conclusions}
We carried out a study of the CO depletion and N$_2$H$^+$ deuteration 
in a sample of representative IM Class 0 protostars.  Our results can be
summarized as follows.

\begin{itemize}
\item We observed the millimeter lines of C$^{18}$O, C$^{17}$O,
N$_2$H$^+$, and N$_2$D$^+$ using the IRAM 30m telescope in a sample of 7 Class 0
and 2 Class I IM stars. We have found a clear evolutionary trend that
differentiates Class 0 from Class I sources. While the emission of the N$_2$H$^+$ 1$\rightarrow$0
peaks towards the star position in Class 0 protostars, it surrounds the
FIR sources in the case of Class I stars. This occurs because the recently formed star
has heated and disrupted the parent core in the case of Class I objects.
The deuterium fractionation, R$_2$, is low, below a few 0.001 in all Class I sources. There is,
however, a wide dispersal in the values of R$_2$ in Class 0 sources ranging from
a few 0.001 to a few 0.01. This  at least two orders of magnitude greater than the
elemental value in the interstellar medium, although a factor of 10-100 lower
than in prestellar clumps.

It is impossible, however, to establish an evolutionary trend among 
Class 0 sources based on simple parameters such as the average CO depletion and
the average N$_2$H$^+$ deuterium fractionation. This stems from the complexity 
of these regions (multiplicity) and the limited angular resolution of our observations, which prevents us from tracing the inner regions of the envelope. Interferometric observations are required to provide a more precise picture of the evolutionary stage of these objects.

\item We used a radiative transfer code to derive the C$^{18}$O, N$_2$H$^+$, and N$_2$D$^+$ radial
abundance profiles in 5 IM Class 0 stars. In particular, we fit the C$^{18}$O 1$\rightarrow$0 maps by assuming that
the C$^{18}$O abundance decreases inwards within the protostellar envelope
until the gas and dust reach the CO evaporation temperature, T$_{ev}$.
Our observational data are better fit with values of T$_{ev}$$\sim$20--25~K,
consistent with the binding energy of 1100~K, which is the
measured in the laboratory for a CO-CO matrix.

\item We determined the chemistry of the protostellar
envelopes using the model by Caselli et al. (2002). 
A  spherical envelope and steady-state chemical model cannot account for the observations.
 We had to introduce modifications to better fit the C$^{18}$O and
N$_2$H$^+$ maps. In particular the CO abundance in the inner envelope seems to be lower than
the canonical value. This could be due to the conversion of CO into CH$_3$OH
on the grain surfaces, the photodissociation of CO by the stellar UV radiation, or even
geometrical effects. Likewise, we have problems fitting the low values of the deuterium fractionation
($\sim$ a few 0.001) measured for some Class 0 IMs. 
Several explanations have been proposed to account for this discrepancy. 
\end{itemize}

\begin{acknowledgements}
This Paper was partially supported by MICINN, within the program
CONSOLIDER INGENIO 2010, under grant  "Molecular
Astrophysics: The Herschel and ALMA Era -- ASTROMOL" (ref.:
CSD2009-00038)
\end{acknowledgements}

\clearpage
\newpage


\begin{appendix} 

\section{}

Below we discuss the details of the modeling for each individual source.

\subsection{Serpens-FIRS~1}

Serpens-FIRS~1, located near the center of the Serpens main core, is
the most luminous object embedded in the cloud. Several continuum studies lead to its classification as a
Class 0 source with a bolometric luminosity estimated to range from 46 L$_\odot$ to 84 L$_\odot$ 
(Harvey et al. 1984; Casali et al. 1993; Hurt \& Barsony 1996; Larsson et al. 2000). The latest estimate of its
luminosity (see Table 4) suggests that Serpens-FIRS~1 is on the low mass/IM borderline.
Serpens-FIRS~1 drives a molecular outflow that is oriented at a position angle of 50$^\circ$
(Rodr\'{\i}guez et al. 1989). C10 modeled this source as a sphere with an outer radius
of 5900~AU, 25.5$"$ at the distance of Serpens. Convolving this size with the observational beams,
one would expect a source diameter of 55$"$, 57$"$, and 53$"$ for the C$^{18}$O 1$\rightarrow$0,
N$_2$H$^+$ 1$\rightarrow$0, and N$_2$D$^+$ 2$\rightarrow$1 maps, respectively. In our maps, the emission is
much more extended ($>$80$"$), showing that the dense molecular cloud significantly contributes to the molecular emission (see also Fig. 6).
To mimic the molecular cloud component, we increased the outer radius in the Crimier et al.\ profile.
We assumed that the radius of the protostellar envelope is 15000~AU, and the dust density and 
temperature smoothly vary between the last point of the Crimier et al.\  profile 
(r=5900~AU, n=5$\times$10$^5$ cm$^{-3}$, T$_d$=13~K) and the  values assumed at r=15000~AU, 
which are n=1$\times$10$^4$~cm$^{-3}$ and T$_d$=10~K. This profile was used for our fitting.

The integrated intensity map of C$^{18}$O cannot be fit
with a constant abundance profile. 
Thus, we decided to assume a step function for the C$^{18}$O abundance:
(i) a constant abundance, X$_{0}$, that is expected to be close to the canonical value (X$_0$=2$\pm$2$\times$10$^{-7}$)
for radii lower than a given radius, R$_0$ and (ii) an abundance, $X_{1}$, that is expected to be $<$X$_0$
for larger radii. Thus defined R$_0$, is the evaporation radius of CO. The values of
X$_0$, R$_0$, and  $X_{1}$ were fit by the model. This approximation was still not sufficient to produce a good fit. We had to add another step, and two new variables, R$_1$ and $X_{2},$ in the C$^{18}$O abundance profile. 
The best fit is shown in Table 5. We have (i) a warm region (R$<$2000~AU) with
an C$^{18}$O abundance $\sim$1.2$\times$10$^{-7}$, (ii) an intermediate layer with a high value for the  
C$^{18}$O depletion, f$_D$$\sim$20, and (iii) an external layer (R$>$6000~AU), which is essentially
the molecular cloud component, in which the C$^{18}$O abundance
is close to the canonical value again.

We followed the same procedure for N$_2$H$^+$. In this case the emission extends to the NE and greatly differs
from spherical symmetry. For this reason we masked the NE quadrant in our fitting (see Fig. 4). Again, we conclude
that the N$_2$H$^+$ abundance has a standard value of a few 10$^{-10}$ in the inner region (R$<$~3000~AU) and in the
molecular cloud, but decreases by at least a factor of 10 in the region between them. 

Serpens-FIRS~1 is one of the three sources where we were able to fit the N$_2$D$^+$ 2$\rightarrow$1 emission.
It is clear from the map morphology that the N$_2$D$^+$ emission is dominated by the molecular
cloud component. Similar to the case of N$_2$H$^+$, we masked the NE quadrant in our fitting. 
We found X(N$_2$D$^+$)=1.9$\times$10$^{-11}$ in the molecular cloud, and obtained an upper limit
of X(N$_2$D$^+$)$<$2.0$\times$10$^{-12}$ for the protostellar core. Thus we have a deuterium fractionation of
$\sim$0.01 in the molecular cloud, and at least an order of magnitude lower in the core (see Fig. 7).

\subsection{Cep~E}

Cep E-mm was cataloged as a Class~0 protostar by Lefloch et al. (1996).
Cep E-mm was observed with IRAM 30m (Lefloch et al. 1996; Chini et al. 2001), SCUBA (Chini et al. 2001), 
ISO (Froebrich et al. 2003), and Spitzer (Noriega-Crespo et al. 2005). All these
studies confirm the Class 0 status of Cep E-mm and constrain the source total mass and
bolometric luminosity in the range of 7-25~M$_\odot$ and 80-120~L$_\odot$, respectively. A bipolar molecular outflow, 
first reported by Fukui et al. (1989),
is associated with Cep~E-mm. The H$_2$ and [FeII] study by
Eisloffel et al. (1996) shows a quadrupolar outflow morphology suggesting that the driving source
is a binary.

C10 modeled this source as a sphere with an outer radius
of 35800~AU, 49$"$ at the distance of Cepheus (see Table 4). Convolving this size with the observational beams,
one would expect a source diameter of $\sim$108$"$, $\sim$110$"$, and 100$"$ for the C$^{18}$O 1$\rightarrow$0,
N$_2$H$^+$ 1$\rightarrow$0, and N$_2$D$^+$ 2$\rightarrow$1 maps, respectively. These sizes agree with those found in our maps.  Of our sources, Cep E has the largest
envelope diameter versus HPBW ratio and thus the best spatial sampling of the varying chemical conditions in
its protostellar envelope (see Table 5).

In this source, we fit the emission
with a constant and close-to-standard abundance of 6$\times$10$^{-8}$ for radii below 3500~AU, and a  
power-law variation of the C$^{18}$O abundance for larger radii (see Table 5). The high value of f$_D$, $\sim$10, 
would occur close to R=3500~AU. The same kind of profile was fit for the N$_2$H$^+$ abundance. In this case,
the quality of the N$_2$D$^+$ 2$\rightarrow$1 line was not good enough to fit the abundance profile. We
derived an averaged abundance across the envelope of $\sim$7$\times$10$^{-12}$ by fitting the spectrum observed towards
the center position. 

\subsection{IC~1396~N}

IC~1396~N is the globule associated with IRAS~21391+5802. It has strong submillimeter and millimeter 
continuum emission (Wilking et al. 1993; Sugitani
et al. 2000; Codella et al. 2001), high density gas (Serabyn et al. 1993; Cesaroni et al. 1999; Codella
et al. 2001; Beltr\'an et al. 2004), and water maser emission (Felli et al. 1992; Tofani et al. 1995;
Patel et al. 2000). IC~1396~N is thus an active site of star formation. 
Using BIMA interferometric millimeter observations, Beltr\'an et al. (2002) detected three
sources (BIMA 1, BIMA 2, and BIMA 3) deeply embedded within the globule. 
Among the three, BIMA 2 has the strongest millimeter emission and is associated with 
an energetic E-W bipolar outflow (Codella et al. 2001
Beltr\'an 2002, 2004). Recent studies by Neri et al. (2007) and
Fuente et al. (2007) using the Plateau de Bure interferometer show that BIMA~2 is itself 
a protocluster composed of 3 cores.

C10 modeled BIMA~2 as a sphere with an outer radius
of 29600~AU, 39$"$ at the distance of IC~1396~N (see Table 4). Together with Cep E-mm, it is
our best-sampled protostellar envelope (large envelope diameter to HPBW ratio). Interferometric observations published by Fuente et al. (2009) reveal that the outflow has already eroded a large biconical cavity in the molecular cloud. This is very likely the reason for the scarce C$^{18}$O 1$\rightarrow$0 and N$_2$H$^+$ 1$\rightarrow$0 emission to the east of the source. We have therefore masked this region to have a more accurate description of the toroidal envelope (see Fig. 4). This masking 
does not affect our results significantly (more than a factor of 2 in the abundances).

Similarly to Cep-E, the C$^{18}$O emission was better fit
with a constant and close-to-standard abundance of 6$\times$10$^{-8}$ in the inner region (R$<$5500~AU) and a  
power-law variation of the C$^{18}$O abundance for larger radii (see Table 5). The highest value of f$_D$ 
is $\sim$5, and it would occur close to R=5500~AU. The N$_2$H$^+$ emission was better fit with a two-step function,
X$_{N_2H^+}$=4.2$\times$10$^{-10}$ for R$<$10000~AU, 1.8$\times$10$^{-10}$ for R$>$15000~AU, and  1.0$\times$10$^{-11}$ in between.
The last step could be due to the vicinity of the sources BIMA 3 and BIMA 2 that heat the outer part
of the envelope. These sources are not considered
in the (n-T) fit by C10. In IC~1396~N, we have not detected the  N$_2$D$^+$ 2$\rightarrow$1 line.
We derived an upper limit to the N$_2$D$^+$ abundance of $<$7$\times$10$^{-12}$.
The lower value of f$_D$ and the non-detection of N$_2$D$^+$ is consistent with this source being a warmer
and more evolved object.

\subsection{CB3}

The CB3 Bok globule is located at $\sim$2.5~kpc (Launhardt \& Henning 1997; Wang et al. 1995).
CB3-mm is the brightest millimeter source
of the globule, first detected by Launhardt \& Henning (1997) and subsequently observed in the
sub-millimeter by Huard et al. (2000). Yun \& Clemens (1994) detected a molecular bipolar outflow in CO, elongated in 
the NE-SW direction, associated with H$_2$O masers (de Gregorio-Monsalvo
et al. 2006). This outflow has been mapped in various molecular lines by Codella \& Bachiller
(1999), who concluded that it originates from CB3-mm. The same authors concluded that CB3-mm is probably a Class 0 source. 

CB3 is different from all the other sources in our sample. In this object,
 the 24~$\mu$m sources are not spatially coincident with the 850~$\mu$m emission peak. CB3-mm hosts
two 24~$\mu$m sources separated by $\sim$12$"$ and neither of them is spatially coincident with the column density peak. The 
column density peak, better traced by the 850~$\mu$m continuum emission, is located in between and 
almost equidistant from the two 24~$\mu$m sources (see Fig. 4).
C10 modeled the SED of this source by adding up the flux of the two 24~$\mu$m sources.
They fit the spatial distribution of the 450~$\mu$m and 850~$\mu$m maps as a sphere with an outer radius
of 103000~AU (i.e. 41$"$) (see Table 4). The radial density power law in this
source is steeper, $\sim$2, than in the others, with low densities ($\sim$a few 10$^3$ cm$^{-3}$) in the outer envelope, 
$>$\ 50000~AU. This means that, although the protostellar envelope is
very large, most of the emission comes from the inner 50000~AU.

We fit the C$^{18}$O emission with a step function.
The abundance is constant at 1.3$\times$10$^{-7}$ for radii less than 25000~AU,  decreases
to  $<$3$\times$10$^{-8}$ ($f_D$$>$3), and then increases to 9.0$\times$10$^{-7}$ for R$>$60000~AU (see Table 5).
The large C$^{18}$O abundance at large radii is very likely caused by the modeled extended low-density envelope providing a poor approximation. More likely, there are dense clumps immersed in 
a lower density cloud. The N$_2$H$^+$ emission is fit with a constant abundance of X$_{N_2H^+}$=6.5$\times$10$^{-10}$. 
The N$_2$D$^+$ emission is fit with a constant abundance of X$_{N_2D^+}$=3.0$\times$10$^{-11}$,
implying an average [N$_2$D$^+$]/[N$_2$H$^+$]=0.05.

\subsection{NGC~7129--FIRS~2}

NGC~7129 is a reflection nebula located in a complex and active star-forming site at a distance of
1250~pc (Hartigan \& Lada 1985; Miranda et al. 1993).
NGC~7129~FIRS~2 is not detected at optical or near-infrared wavelengths. Its position 
is spatially coincident with a $^{13}$CO column
density peak (Bechis et al. 1978) and a high-density NH$_3$ cloudlet (Guesten \& Marcaide 1986),
and it is close to an H$_2$O maser (Rodr\'{\i}guez et al. 1980). NGC7129--FIRS~2 has been classified as
a Class 0 IM protostar by Eiroa et al. (1998), who carried out a multi-wavelength study of the continuum
emission from 25 to 2000~$\mu$m. Edwards \& Snell (1983) detected a bipolar CO
outflow associated with FIRS~2. The interferometric study by Fuente et al. (2001) reveals that
this outflow presents a quadrupolar morphology. In fact, the outflow seems to be the superposition of two flows, 
FIRS~2-out~1 and FIRS~2-out~2, likely associated with FIRS~2 and a more evolved star (FIRS~2-IR), respectively. 
Fuente et al. (2005a,b) carried out a complete chemical study of FIRS 2 providing the first detection
of hot core in an IM Class 0. Based on all these studies, FIRS 2 is considered the youngest IM object
known at present.

C10 modeled NGC~7129--FIRS~2 as a sphere with an outer radius
of 18600~AU, 15$"$ at the distance of NGC~7129 (see Table 4). 
Since the source is very compact and barely resolved in the maps of the
C$^{18}$O 1$\rightarrow$0 and N$_2$H$^+$ 1$\rightarrow$0 lines, we fit only an average abundance in
for these species. The average CO depletion factor is f$_D$=3, consistent with the depletion factor found by Fuente et al. (2005a)
based on H$^{13}$CO$^+$ observations. The protostellar envelope is, however, resolved at the frequency of the N$_2$D$^+$ 3$\rightarrow$2
line. Since previous interferometric observations published by Fuente et al. (2005a) reveal that the
N$_2$D$^+$ emission is absent towards the hot core, we fitted the N$_2$D$^+$ 3$\rightarrow$2 emission
assuming that it arises in a ring. The best fit is for a ring with inner radius of 9000~AU and an outer radius of 11000~AU. Within the
ring the N$_2$D$^+$ abundance is $\sim$4.5$\times$10$^{-11}$, and the deuterium fractionation, $\sim$0.1, is the typical
value found in prestellar cores.

\end{appendix}

\end{document}